\documentclass[manuscript]{aastex}

\usepackage{natbib}

\slugcomment{Not to appear in Nonlearned J., 45.}

\shorttitle{Observations of Post-flare Loop System}
\shortauthors{Srivastava \& Murawski}

\begin{document}


\title{Observations of Post-flare Plasma Dynamics during an M1.0 Flare in AR11093 by SDO/AIA*}


\author{Abhishek K. Srivastava\altaffilmark{1,2}}
\affil{Aryabhatta Research Institute of Observational Sciences (ARIES), Manora Peak, Nainital-263 129, India}

\author{K. Murawski\altaffilmark{2}}
\affil{Group of Astrophysics,UMCS, ul. Radziszewskiego 10, 20-031 Lublin, Poland}
\email{kmur@kft.umcs.lublin.pl}


\altaffiltext{1}{Aryabhatta Research Institute of Observational Sciences (ARIES), Manora Peak, Nainital-263 129, India.}
\altaffiltext{2}{Group of Astrophysics,UMCS, ul. Radziszewskiego 10, 20-031 Lublin, Poland.}
\altaffiltext{*}{AKS dedicates the present research work and all his solar research to his grandmother 
Gulab Devi (8 January 1919-16 September 2011).}


\begin{abstract}
We observe the motion of cool and hot plasma in a multi-stranded post flare 
loop system that evolved in the decay phase of a two ribbon M1.0 class flare 
in AR 11093 on 7 August 2010 using SDO/AIA 304 \AA\ and 171 \AA\ filters. 
The moving intensity feature and its reflected counterpart are observed in the 
loop system at multi-temperature.
The observed hot counterpart of the plasma 
that probably envelopes the cool confined plasma, moves comparatively faster
($\sim$34 km s$^{-1}$) to the 
later (29 km s$^{-1}$) in form of the spreaded intensity feature. The propagating plasma and 
intensity reflect from the region of another footpoint of the loop. 
The subsonic speed of the moving 
plasma and associated intensity feature may be most likely evolved in
the post flare loop system through impulsive flare heating processes. 
Complementing our observations 
of moving multi-temperature intensity features in the post flare loop system and its
reflection, we also attempt to
solve two-dimensional ideal magnetohydrodynamic equations 
numerically using the VAL-IIIC atmosphere as an initial condition 
to simulate the observed plasma dynamics. We consider a localized 
thermal pulse impulsively generated near one footpoint of the loop 
system during the flare processes, which is launched along the magnetic field lines  
at the solar chromosphere. The pulse steepens into a slow shock at higher altitudes 
while moving along this loop system, which triggers plasma perturbations 
that closely exhibit the observed plasma dynamics. 
\end{abstract}





\section{Introduction}
Post flare loops are the dynamic and transient loop systems that evolve in flaring active regions
, and exhibit strong localized heating and cooling effects,
as well as emissions in various parts of the electromagnetic spectrum (e.g., HXR, SXR,
EUV, H$_{\alpha}$) during the gradual reconnection process occurred in the flare activity \citep{
Hara98, Reevs02, Li09b}. Recently, \citet{Cheng10} have observed a 
re-flaring event in the evolved post flare loop system triggered by
the magnetic flux emergence. \citet{Li09a} have observed the brightness propagation
in post flare loop (PFLs) systems. They have investigated that the dynamics of
these brightened PFLs were along the neutral lines
and separation of the flare ribbons were perpendicular to the neutral lines.
Their observations indicate that the footpoints of the initially brightened PFLs were 
always associated with the change of the photospheric magnetic fields, and 
support the general 3-D model of the magnetic reconnection \citep{shio05}.
\citet{War99} have observed impulsive footpoint brightening in a two ribbon flare followed by the formation of high-temperature plasma in the corona and then cooler post flare loops below the hot plasma, which also supports the standard reconnection model of solar flares.
There are several mechanisms to understand the heating of post flare loops in which
the most likely mechanism of their heating and rise-up may be attributed to slow magnetohydrodynamic shock waves \citep{Car82}. 
Recently, \citet{Fern09} have numerically simulated the internal plasma dynamics
in such post flare loops, and concluded that the high-speed downflow perturbations, 
usually interpreted as slow magnetoacoustic waves before, 
could now be interpreted as slow magnetoacoustic shock waves.
\citet{Dana09} have also reported the field lines shortening in post flare loop system
and inferred the localized, three-dimensional reconnection to heat the plasma. The shortening progresses 
away from the reconnection location with an Alfv\'en speed and release magnetic energy and 
generate parallel compressive flows. 

The post-flare loops are not only the most dynamic magnetic flux tubes associated with heating, cooling, and 
mass flows, but also the excellent waveguides for the excitation of various
types of magnetohydrodynamic waves and oscillations. \citet{Eho07} have observed 
the simultaneous presence of propagating disturbances and multiple harmonics 
of the fast magnetoacoustic kink oscillations in the transition region post flare loops
in their cooling phase. \citet{Sri08} have also observed the multiple harmonics of the
sausage oscillations in the cool post flare loops along with a strong downflow
of the plasma. Recently, \citet{Val11} have reported the propagating disturbances 
in the loop system associated with two-ribbon flare as a signature of slow
magnetoacoustic waves. Moreover, \citet{Kry04} have found that two types of waves 
are generated in post flare loops due to the growth of instabilities, e.g., 
the kinetic Alfv\'en-like waves and  magnetoacoustic waves.
The observations of waves and oscillations in the 
highly dynamic post flare loops provide compelling motivation to diagnose their
crucial plasma properties using principle of MHD seismology to understand 
its dynamics \citep[and references cited there]{And09, Asc09, Mac10, Mac11, Abhi10}.

In the present paper, we present the recent SDO/AIA observations
of the moving intensity features seen in a multi-temperature
plasma in a multi-stranded post flare loop system whose footpoints
are anchored in the two opposite magnetic polarities at two
flare ribbons. We do not aim to study the large-scale 
post flare loop dynamics during whole long duration decay phase 
of the associated M-class flare in AR11093 on 7 August 2010. However, we do aim to analyse the
most probable cause of the internal dynamics of the moving intensity feature
and its reflection in the loop system during 15 minutes (18:44 UT-18:58 UT) 
when it was almost in equilibrium. 
In an attempt to understand the dynamics and plasma conditions of the observed
post flare loop system that is rather complex, we 
numerically model the plasma dynamics by solving
2-D ideal MHD equations using the VAL III C 
thermodynamic structure as an initial condition  
by triggerring a thermal pulse near one footpoint
of the simulated loop system.
Our numerical model provides only a rough guide for 
the observed complex plasma dynamics as it 
does not include thermal condution and radiative cooling 
. In the present paper we mainly focus on the unique
observations of the post flare plasma dynamics during an
M-class flare, but we also briefly present the 
supportive results of munerical modeling that 
approximately exhibit the observed internal plasma dynamics.
In Sec. 2,  we present observations and results. We describe the
numerical models in Sec. 3. In Sec. 4, we report the
results of numerical simulations. The discussion and conclusions are
given in the last section.

\section{Observations and Results}
We use the time series data of post flare phase of a two ribbon 
M1.0 class solar flare in AR11093 on 7 August 2010 as observed in a 304 \AA \ 
and 171 \AA\ filters of the Atmospheric 
Assembly Imager (AIA) onboard Solar Dynamics Observatory (SDO)
on 7 August 2010. Although the AR 11093 evolution and M1.0 class flare are observed by all
the SDO/AIA filters as well as other instruments (e.g., STEREO-B, SoHO/EIT, MDI etc),
we present the high-resolution SDO/AIA observations of the evolution of
a post flare loop system during 18:44 UT-18:58 UT in just two filters 
304 \AA\ and 171 \AA\ in order to study the unique post flare multi-temperature internal plasma dynamics.
These two filters represent the plasma emissions formed respectively
at temperatures $\sim$0.1 MK and $\sim$1.0 MK.
The SDO/AIA has a typical resolution of 0.6$"$ per pixel
and the highest cadence of 12 s, and it observes the
full solar disk in three UV (1600 \AA, 1700\AA, 4500 \AA)
and seven EUV (171 \AA, 193 \AA, 211 \AA, 94 \AA, 304 \AA, 335 \AA, 131 \AA) lines \citep{Lem11}. 
In Fig.1, the enlarged field-of-view of the two-ribbons and post flare phase of an M1.0 class 
flare in AR11093 are shown at 18:51 UT on 7 August 2010 as observed 
by SDO/AIA 304 \AA\ (left-panel) and 171 \AA\ (right-panel). The SoHO/MDI
magnetic polarities are over plotted above these snap-shots that clearly reveal
the opposite field distributions (+ve with yellow and -ve with magenta) over the
two flare ribbons. The MDI contour levels are $\pm$800 G of the magnetic field strength.
The time series has been obtained by the SSW cutout service at LMSAL, USA, which
is corrected for the flat-field and spikes. We run aia\_prep subroutine 
of SSW IDL for further calibration and cleaning of the time
series data.

As per Hinode/EIS flare catalog \footnote{http://msslxr.mssl.ucl.ac.uk:8080/SolarB/eisflarecat.jsp}
,the M1.0 class flare begins in AR 11093 (N11$^{o}$ E34$^{o}$) on
$\sim$17:55 UT, peaks at $\sim$18:24 UT, and subsided 
on $\sim$18:47 UT. However, as per GOES X-ray 
flux measurements, its decay phase is 
extended upto $\sim$20:50 UT making it almost a long duration event (LDE).
A part of huge and twisted filament that was aligned along 
the magnetic neutral line, rose and erupted. 
The activated flux-rope and its reconnection with the overlying coronal loops may 
most likely begin
the flare energy release process. 
The detailed morphological study of this M1.0 class flare and associated 
CME has been recently reported by \cite{Reddy11}. The transient brightening of 
the two ribbons occurred due to the precipitation of energetic particles towards 
the loop footpoints that participated in gradual reconnection process in the corona. Therefore, the heating of the
low atmospheric plasma along the two ribbons uploads the mass and forms
a new set of the post flare loops in the decay phase of the flare (cf., Fig. 1).
In this paper, we aim only to study the short duration internal plasma dynamics 
of a post flare loop system during M1.0 class flare in the observations.

In Fig. 2, we observe a very unique evolution of multi-stranded post flare loop system on
18:44 UT. A moving intensity feature is clearly evident in the loop system on 18:46 UT,
which propagates down towards southward footpoint of the 
loop system. It gradually brightens and moves downward till 18:52 UT, and thereafter,
it still propagates downward with a subsequent decrease in its intensity. It
is not confined in
one strand, but seems to propagate in the multiple strands of the post flare loop system.
The southward positive polarity footpoint (yellow MDI contours) and northward negative polarity
footpoint (magenta MDI contours) of the loop system are plotted in the first and last snapshots
of Fig. 2. The contour levels are with $\pm{800}$ G 
magnetic field strength. The multiple strands of the loop system are clearly anchored in these two foot points.
We estimate the approximate lower bound velocity of this moving intensity feature as
$\sim$29 km s$^{-1}$. This is a sub-sonic speed as the local sound speed at the formation
temperature (T$_{f}$$\sim$1$\times$$10^{5}$ K) of He II ion that emits 304 \AA\ , is around 46 km s$^{-1}$.
The interesting scenario became evident in form of the antipropagation of 
moving intensity during 18:57-18:58 UT, which indicates the reflection of this low temperature counterpart of the plasma 
near the southward footpoint of the post flare loop system.
However, it should be noted that the moving intensity feature could not exactly reach 
to the southward footpoint of loop, but it started reflecting back from its nearby
location.

In Fig. 3, we also observe this very unique evolution of multi-stranded post flare loop system at coronal 
temperature ($\sim$1.0 MK) in 171 \AA\ filter of SDO/AIA. The bright post flare loop system is clearly 
appeared on 18:44 UT. A brightness associated with high temperature counterpart of the plasma, however, is 
spreaded over the larger regions in the post flare
loop system and clearly evident on 18:45 UT.
This means that the high temperature counterpart of the plasma envelopes its low temperature counterpart
and form the co-spatial multitemperature plasma blob.
The high temperature counterpart of the moving intensity propagates down towards southward footpoint of the 
loop system. It gradually brightens and moves downward till 18:51 UT, and thereafter
it still propagates downward with decreasing intensity. It 
is clearly visible in the coronal image on 18:51 UT that the near footpoints areas of these multiple strands
are merged with each other and forming a multiple stranded post flare loop system.
The brightness is not confined in
one strand, but it seems to propagate in multiple strands of the post flare loop system.
We estimate the approximate lower bound velocity of this moving intensity feature as
$\sim$34 km s$^{-1}$. This is a sub-sonic speed as the local sound speed at the formation
temperature (T$_{f}\sim$1.0 MK) of Fe IX ion that emits 171 \AA , is $\sim$147 km s$^{-1}$.
The interesting scenario became evident again in form of the antipropagation of 
the moving intensity during 18:51-18:58 UT in various parts of the loop system, which indicates the reflection 
of heated plasma also from nearby region of the southward footpoint.
The hot plasma moves slightly faster compared to its 
cooler counterpart as observed in Figs. 2 and 3, however, both
are subsonic in nature. Therefore,
the hot plasma is well spread over the larger area, while the
low temperature plasma is confined to a smaller region. However, they move approximately co-spatially
in a mixed state where hot plasma envelopes the cool confined plasma.
This means that the moving intensity feature is made 
by cool and hot components of the plasma, which moves towards southern footpoint of the loop system
and reflects back.
However, it should be noted in the case of hot plasma counterpart as visible in coronal line 
that the moving intensity feature also could not reach exactly  
to the southward footpoint of the loop, but it started reflecting back from its nearby
location.

\section{A Numerical Model of Plasma Dynamics}
%
%
%
To model the observed plasma dynamics in the
post flare loop system, we consider a gravitationally-stratified solar atmosphere
which is described by
the ideal 2D
MHD equations:
\begin{equation}
\label{eq:MHD_rho}
{{\partial \varrho}\over {\partial t}}+\nabla \cdot (\varrho{\bf V})=0\, ,
\end{equation}
\\
\begin{equation}
\label{eq:MHD_V}
\varrho{{\partial {\bf V}}\over {\partial t}}+ \varrho\left ({\bf V}\cdot \nabla\right ){\bf V} =
-\nabla p+ \frac{1}{\mu}(\nabla\times{\bf B})\times{\bf B} +\varrho{\bf g}\, ,
\end{equation}
\\
\begin{equation}
\label{eq:MHD_p}
{\partial p\over \partial t} + \nabla\cdot (p{\bf V}) = (1-\gamma)p \nabla \cdot {\bf V}\, ,
\hspace{3mm}
p = \frac{k_{\rm B}}{m} \varrho T\, ,
\end{equation}
\\
\begin{equation}
\label{eq:MHD_B}
{{\partial {\bf B}}\over {\partial t}}= \nabla \times ({\bf V}\times{\bf B})\, ,
\hspace{3mm}
\nabla\cdot{\bf B} = 0\, .
\end{equation}
%
Here ${\varrho}$ is mass density, ${\bf V}$ is flow velocity,
${\bf B}$ is the magnetic field, $p$ is gas pressure, $T$ is temperature,
$\gamma=5/3$ is the adiabatic index, ${\bf g}=(0,-g)$ is gravitational acceleration of
its value $g=274$ m s$^{-2}$,
$m$ is mean particle mass and $k_{\rm B}$ is the Boltzmann's constant.
It should be noted that our model does not include either the 
electron thermal conduction or radiative cooling for simplicity, which may have
significant effect on the observed plasma dynamics. These terms
are also known to dominant the energy equations
in the corona and transition region. Nevertheless, we
expect the velocity response of the plasma to an impulsive burst of heating
to behave qualitatively correctly.
\subsection {Equilibrium Configuration}
%
%
%
We assume that the solar atmosphere is in static equilibrium (${\bf V}_{\rm e}=0$) with a force-free magnetic field,
\begin{equation}
\label{eq:B}
(\nabla\times{\bf B}_{\rm e})\times{\bf B}_{\rm e} = 0\, ,
\end{equation}
%
the pressure gradient is balanced by the gravity force,
\begin{equation}
\label{eq:p}
-\nabla p_{\rm e} + \varrho_{\rm e} {\bf g} = 0\, .
\end{equation}
Here the subscript $_{\rm e}$ corresponds to equilibrium quantities.

Using the ideal gas law and the $y$-component of hydrostatic
pressure balance indicated by Eq.~(\ref{eq:p}), we express
equilibrium gas pressure and mass density as
\begin{equation}
\label{eq:pres}
p_{\rm e}(y)=p_{\rm 0}~{\rm exp}\left[ -\int_{y_{\rm r}}^{y}\frac{dy^{'}}{\Lambda (y^{'})} \right]\, ,\hspace{3mm}
\label{eq:eq_rho}
\varrho_{\rm e} (y)=\frac{p_{\rm e}(y)}{g \Lambda(y)}\, .
\end{equation}
Here
\begin{equation}
\Lambda(y) = k_{\rm B} T_{\rm e}(y)/(mg)
\end{equation}
is the pressure scale-height, and $p_{\rm 0}$ denotes the gas
pressure at the reference level that we choose in the solar corona at $y_{\rm r}=10$ Mm.

We adopt
an
equilibrium temperature profile $T_{\rm e}(z)$ for the solar atmosphere
that is close to the VAL-C atmospheric model of Vernazza et al. (1981). It is smoothly extended into the corona.
Then with the use of Eq.~(\ref{eq:pres})
we obtain the corresponding gas pressure and mass density profiles.

We assume that the initial magnetic field satisfies a current-free condition,
%
$\nabla \times \vec B_{\rm e}=0$, and it is specified by the magnetic flux function, $A$,
such that
$$
\vec B_{\rm e}=\nabla \times (A\hat {\bf z})\, .
$$
%
We set an arcade magnetic field by choosing
%
\begin{equation}
A(x,y) = B_{\rm 0}{\Lambda}_{\rm B}\cos{(x/{\Lambda}_{\rm B})} {\rm exp}[-(y-y_{\rm r})/{\Lambda}_{\rm B}]\, .
\end{equation}

%

%
%
%

Here, $B_{\rm 0}$ is the magnetic field at $y=y_{\rm r}$, 
which is set by requiring that at $y=y_{\rm r}$ Alfv\'en speed $c_{\rm A}=B_{\rm 0}/\sqrt{\mu \varrho_{\rm e}(y_{\rm r})}$ is 
$10$ times larger than sound speed $c_{\rm s}=\sqrt{\gamma p_{\rm e}(y_{\rm r})/\varrho_{\rm e}(y_{\rm r})}$. 
This requirement specifies $B_{\rm 0}\simeq 0.0013$ Tesla. 
The magnetic scale-height is
\begin{equation}
{\Lambda}_{\rm B}=2L/\pi\, .
\end{equation}
We use $L=75$ Mm. 
%
%

%
%
\subsection{Perturbations}
%
We initially perturb
the above equilibrium impulsively by a Gaussian pulse in the
gas pressure $p$, 
viz.,
\begin{equation}\label{eq:perturb}
p(x,y,t=0) = A_{\rm p} \exp\left[ -\frac{(x-x_{\rm 0})^2}{{w_{\rm x}}^2}-\frac{(y-y_{\rm 0})^2}{{w_{\rm y}}^2}\right]\, .
\end{equation}
Here $A_{\rm P}$ is the amplitude of the pulse, $(x_{\rm 0},y_{\rm 0})$ is its initial position and
$(w_{\rm x}, w_{\rm y})$ denotes its widths along the x and y directions. 
We set and hold fixed $x_{\rm 0}$=60 Mm, $y_{\rm 0}$=1.75 Mm, $w_{\rm x}$=2.5 Mm, $w_{\rm y}$=0.4 Mm
, and $A_{\rm p}$=30 $p_{\rm e}$ at $y$=$y_{\rm 0}$.
%

%
\section{Results of Numerical Simulations}
Equations (\ref{eq:MHD_rho})-(\ref{eq:MHD_B}) are solved numerically using the code FLASH
\citep{Lee09}. This code implements a second-order unsplit Godunov solver 
with various slope 
limiters and Riemann solvers, as well as adaptive mesh refinement (AMR).
We set the simulation box 
of $(0,70)\, {\rm Mm} \times (0,40)\, {\rm Mm}$ along the $x$- and $y$-directions 
and impose fixed in time all plasma quantities at all four boundaries of the simulation region. 
In all our studies we use AMR grid with a minimum (maximum) level of 
refinement set to $4$ ($7$). The refinement strategy is based on 
controlling numerical errors in 
mass density, which results in an excellent resolution of steep spatial profiles and
greatly reduces numerical diffusion at these locations.

Fig.4 displays the spatial profiles of plasma temperature (colour maps) and velocity (arrows) 
resulting from the initial pressure pulse of Eq.~(\ref{eq:perturb}), which splits into counter-propagating parts. The part which propagates downwards 
becomes reflected from the dense plasma layers at the photospheric region. This reflected part lags behind the originally upward propagating 
signal that becomes a slow shock. The applied initial pressure pulse with its amplitude 30 times larger
than the same for ambient plasma, was launched at the right foot-point
of a loop system.
As the plasma is initially pushed upwards, an under-pressure region is created below the initial pulse. This under-pressure region
sucks up 
comparatively less heated plasma, which lags behind the high temperature shock front. 
As a result, the pressure gradient works against the gravity and forces 
the material to penetrate into the higher parts of the loop system in form of rarefaction wave. 
It should be noted that the observed post flare loop systems were not visible initially, and they 
became visible after filling up of the plasma during the flare process. In the observations, we 
choose one multi-stranded post flare loop system whose footpoints were separated
approximately by 50 Mm, and the width was $\sim$7 Mm. We set our simulated 
loop system approximately with the same spatial scale of the selected post flare loop system in the observations.
In the simulation of internal plasma dynamics of post flare loop system, the snapshot at $t=100$ s will be a reference in which the
plasma is lifting up from its one footpoint with the launch of pressure pulse. 
A similar scenario is also evident in 18:44 UT images of the observations (cf., Figs 2-3).
It should also be noted that the cool counterpart of the plasma may be enveloped by
its hot counterpart in the simulations similar to the observations. However, initially it may not be realized in the more
ideal simulation domain. It is even also difficult to distinguish hot and cool plasma
in the observations exactly due to
the projection effect as well as dominant coronal emissions in the loop system.
However, at $t=400-500$ s, few hot plasma threads are flowing laminar to the cool plasma that may support the 
observational scenario upto some extent. However,
we can not exactly mimic the complex real observational scenario of the presence of multi-temperature plasma
in the simulation.
At $t=400$ s and later, the slow shock with high temperature front hits at the left footpoint of the loop system 
and reflects back from the dense plasma near this footpoint. 
The moving plasma interacts with the
reflected shock (cf., t=400-800 s simulation snapshots). Therefore,
the moving plasma does not reach exactly 
at the left footpoint and starts propagating
in the opposite direction. This is clearly evident in the simulation,
and this scenario matches well with the observations as displayed in 
the snapshots at 18:58 UT in Figs 2-3. In the simulation, the same scenario
became well evident  around $t$=800 s when the slow reflected shock
interacts with the moving plasma and causes its antipropagation.
It should be noted that the moving plasma is also sub-sonic 
in the numerical simulation that matches well with the observed
moving intensity feature.

Due to the interaction of reflected shock and the downward moving plasma  
, the  plasma
becomes much compressed near the left foot-point region. Therefore, at this location, the
high pressure region does
not allow the moving plasma to fall down towards the left foot-point completely due to the
gravity. A similar situation is evident in the observations when a mixture of plasma containing both heated and cool material
again starts propagating towards the northward footpoint in the post flare loop system (cf., Figs 2-3, 18:57-18:58 UT).
In conclusion, although the real observational situation 
exhibits the complex dynamics of a multi-temperature plasma, our numerical
results approximately reproduce the observed post flare plasma dynamics
despite our use of a simplified model. 


\section{Discussions and Conclusion}\label{SECT:DISS}
In the present work, we observe the unique dynamics of cool and hot plasma in a multi-stranded post flare 
loop system that evolved in the decay phase of a two ribbon M1.0 class flare 
in AR 11093 on 7 August 2010 using SDO/AIA 304 \AA\ and 171 \AA\ filters. 
The moving intensity feature and its reflected counterpart are observed in the 
loop system at multi-temperatures.
The observed hot counterpart of the plasma 
that probably envelopes the cool confined plasma, moves comparatively faster
(34 km s$^{-1}$) to the 
later component (29 km s$^{-1}$) in form of the spreaded intensity feature. The propagating plasma and 
intensity reflect from another footpoint of the loop. 
The subsonic speed of the moving 
plasma and associated intensity feature may be most likely evolved in
the post flare loop system impulsively by the flare heating processes
of the loop footpoints.

Recently, \citet{Val11} have demonstrated that sub-sonic disturbances propagating along the axis of post flare arcades in two-ribbon flares can be interpreted in terms of slow magnetoacoustic waves. These waves can propagate across the magnetic field, parallel to the magnetic neutral line, because of the wave-guiding effect due to the reflection from the footpoints. Our observed moving intensity feature is also subsonic but propagates along the loop strands  rather than along the magnetic neutral line. Using TRACE observations of various flaring events, \citet{Li09a} have statistically studied the dynamics of the propagating brightness features in the post flare loop arcades. They have found that the brightness propagations move with the speeds in the range of 3-39 km s$^{-1}$. Our observed speeds of the hot and cool counterparts of the plasma also lie in the same range as observed by \citet{Li09a}, which were subsonic motions. They have observed three kind of internal motions in the post flare loop systems, e.g., (i) the excitation of propagating brightness near loop apex that moves towards both the footpoints, (ii) the triggering of perturbation at one end that propagates towards other end, and (iii) multi-spatial excitation and propagation of the brightening in post flare loop systems (PFLs). They have concluded a general 3-D reconnection scenario for the generation of their observed PFL plasma dynamics. They have also found that the propagation of brightness in post flare loops are coupled with the ribbon separation.  However, our observed unique moving intensity feature was the propagation of multi-temperature subsonic plasma starting near one footpoint and then its reflection from another footpoint of the post flare loop system. Moreover, in our case, the
post flare loop system and the flare ribbons were almost in equilibrium for chosen duration (18:44 UT--18:59 UT) of 15 min. Therefore, we could track the internal plasma dynamics of loop system to compare with the numerical simulation. The two ribbon M1.0 class flare has occurred in a classical way as per the standard reconnection model \citep[e.g,][]{shio05}. The part of filament which was located along the neutral line in between both the ribbons, most likely activated and subsequently reconnected with the overlying coronal field lines that triggered an M1.0 class flare, followed by a long duration post flare phase. 
However, we observe unique and short duration internal plasma dynamics when the ribbon positions and the post flare loop were almost stable. Therefore, the flare occurred most likely due to standard 3-D multi-stage reconnection event \citep{shio05}, had already subsided for the chosen duration and was not leading much ribbon separation as well as the bulk dynamics of evolved post flare loop systems along these two ribbons. Therefore, we suggest that some local driver generated by impulsive heating due to standard reconnection process leads the observed plasma dynamics in the selected post flare loop  system. The reconnection generated impulsive energy deposition by the energetic particles to the ribbons may lead the density as well as pressure perturbations near the loop footpoints. The triggered thermal purturbations may steepen into a slow shock front moving along the loop threads and generates the rarefraction wave behind, which causes the motion of brightened plasma up as observed by SDO/AIA 304 \AA\ and 171 \AA\ filters.  The interaction of the hot and cool plasma components with the reflected leading shock causes the antipropagation of multi-temperature plasma and generates the complex dynamics of the loop system as observed by SDO/AIA.

We also attempt a numerical simulation to show a general scenario of an observed post flare plasma dynamics internally in a post flare loop system. However, the real dynamics was excited in more complex plasma and magnetic field conditions in the two-ribbon M1.0 class flaring region in AR 11093 on 7 August 2010. We might not model exactly realistically the way in which the post flare loop system was evolved in the real Sun, and therefore its complex plasma dynamics. In fact, the excitation mechanism could work for some time, it could be located at a different place and it could have a different size, pressure perturbation amplitude and distribution. However, in our simulation we trigger the plasma dynamics by a localized pulse in the pressure that is launched below the transition region. In this way we managed to excite the plasma dynamics which qualitatively mimics on average the observations. The small mismatches like the exact time-scales of the motion of plasma and starting of its reflection etc can be matched by tuning of free parameters in the numerical model.

In conclusion we suggest that the initial thermal pulse launched below the transition region is able to trigger a slow shock 
manifested mainly as a plasma flow along the loop strands, which is followed by a brightened moving plasma material. This scenario resembles the observed fine structural dynamics of the post flare loop system. We report first time on the observations of a thermal pulse driven multi-temperature plasma in the post flare loop and provide an approximate theoretical explanation of this phenomenon on the basis of numerical 
simulations we performed. However, further multiwavelength observations should be performed by high-resolution space borne (e.g., SDO, Hinode, STEREO) and complementary ground based observations to shed new light on this kind of unique plasma dynamics. This will also impose a rigid constraint on the stringent simulations of such kind of dynamics in the model solar atmosphere, e.g., the consideration of
more realistic atmosphere by taking into account the thermal conduction and radiative cooling effects. 

\section{Acknowledgments}
We thank reviewer for his/her valuable suggestions that improved the manuscript considerably. We acknowledge the use of the SDO/AIA observations for this study. The data is provided curtesy of NASA/SDO, LMSAL, and the AIA, EVE, and HMI science teams. The FLASH code has been developed by the DOE-supported ASC/Alliance
Center for Astrophysical Thermonuclear Flashes at the University of
Chicago. AKS acknowledges Shobhna Srivastava for the patient encouragement.
KM thanks Kamil Murawski for his helps in the
visualization of the numerical results.

\bibliographystyle{apj} 
\bibliography{reference} 

\begin{thebibliography}{}

\bibitem[\protect\citeauthoryear{{Andries} et~al.}{{Andries}
  et~al.}{2009}]{And09}
{Andries}, J., {van Doorsselaere}, T., {Roberts}, B., {Verth}, G., {Verwichte},
  E.,  \& {Erd{\'e}lyi}, R. 2009, \ssr, 149, 3

\bibitem[\protect\citeauthoryear{{Aschwanden}}{{Aschwanden}}{2009}]{Asc09}
{Aschwanden}, M.~J. 2009, \ssr, 149, 31

\bibitem[\protect\citeauthoryear{{Cargill} \& {Priest}}{{Cargill} \&
  {Priest}}{1982}]{Car82}
{Cargill}, P.~J.,  \& {Priest}, E.~R. 1982, \solphys, 76, 357

\bibitem[\protect\citeauthoryear{{Cheng} et~al.}{{Cheng}
  et~al.}{2010}]{Cheng10}
{Cheng}, X., {Ding}, M.~D., {Guo}, Y., {Zhang}, J., {Jing}, J.,  \&
  {Wiegelmann}, T. 2010, \apjl, 716, L68

\bibitem[\protect\citeauthoryear{{Fern{\'a}ndez} et~al.}{{Fern{\'a}ndez}
  et~al.}{2009}]{Fern09}
{Fern{\'a}ndez}, C.~A., {Costa}, A., {Elaskar}, S.,  \& {Schulz}, W. 2009,
  \mnras, 400, 1821

\bibitem[\protect\citeauthoryear{{Harra-Murnion} et~al.}{{Harra-Murnion}
  et~al.}{1998}]{Hara98}
{Harra-Murnion}, L.~K., et~al. 1998, \aap, 337, 911

\bibitem[\protect\citeauthoryear{{Kryshtal} \& {Gerasimenko}}{{Kryshtal} \&
  {Gerasimenko}}{2004}]{Kry04}
{Kryshtal}, A.~N.,  \& {Gerasimenko}, S.~V. 2004, \aap, 420, 1107

\bibitem[\protect\citeauthoryear{{Lee} \& {Deane}}{{Lee} \&
  {Deane}}{2009}]{Lee09}
{Lee}, D.,  \& {Deane}, A.~E. 2009, Journal of Computational Physics, 228, 952

\bibitem[\protect\citeauthoryear{{Lemen}, {Title}, \& {Akin}}{{Lemen}
  et~al.}{2011}]{Lem11}
{Lemen}, J.~R., {Title}, A.~M.,  \& {Akin}, D. e.~a. 2011, \aap, A8

\bibitem[\protect\citeauthoryear{{Li} \& {Zhang}}{{Li} \&
  {Zhang}}{2009a}]{Li09b}
{Li}, L.,  \& {Zhang}, J. 2009a, \apj, 703, 877

\bibitem[\protect\citeauthoryear{{Li} \& {Zhang}}{{Li} \&
  {Zhang}}{2009b}]{Li09a}
{Li}, L.,  \& {Zhang}, J. 2009b, \apj, 690, 347

\bibitem[\protect\citeauthoryear{{Longcope}, {Guidoni}, \& {Linton}}{{Longcope}
  et~al.}{2009}]{Dana09}
{Longcope}, D.~W., {Guidoni}, S.~E.,  \& {Linton}, M.~G. 2009, \apjl, 690, L18

\bibitem[\protect\citeauthoryear{{Macnamara} \& {Roberts}}{{Macnamara} \&
  {Roberts}}{2010}]{Mac10}
{Macnamara}, C.~K.,  \& {Roberts}, B. 2010, \aap, 515, A41

\bibitem[\protect\citeauthoryear{{Macnamara} \& {Roberts}}{{Macnamara} \&
  {Roberts}}{2011}]{Mac11}
{Macnamara}, C.~K.,  \& {Roberts}, B. 2011, \aap, 526, A75

\bibitem[\protect\citeauthoryear{{Nakariakov} \& {Zimovets}}{{Nakariakov} \&
  {Zimovets}}{2011}]{Val11}
{Nakariakov}, V.~M.,  \& {Zimovets}, I.~V. 2011, \apjl, 730, L27

\bibitem[\protect\citeauthoryear{{O'Shea} et~al.}{{O'Shea}
  et~al.}{2007}]{Eho07}
{O'Shea}, E., {Srivastava}, A.~K., {Doyle}, J.~G.,  \& {Banerjee}, D. 2007,
  \aap, 473, L13

\bibitem[\protect\citeauthoryear{{Reddy}, {Ajor Maurya}, \& {Ambastha}}{{Reddy}
  et~al.}{2011}]{Reddy11}
{Reddy}, V., {Ajor Maurya}, R.,  \& {Ambastha}, A. 2011, ArXiv e-prints

\bibitem[\protect\citeauthoryear{{Reeves} \& {Warren}}{{Reeves} \&
  {Warren}}{2002}]{Reevs02}
{Reeves}, K.~K.,  \& {Warren}, H.~P. 2002, \apj, 578, 590

\bibitem[\protect\citeauthoryear{{Shiota} et~al.}{{Shiota}
  et~al.}{2005}]{shio05}
{Shiota}, D., {Isobe}, H., {Chen}, P.~F., {Yamamoto}, T.~T., {Sakajiri}, T.,
  \& {Shibata}, K. 2005, \apj, 634, 663

\bibitem[\protect\citeauthoryear{{Srivastava}}{{Srivastava}}{2010}]{Abhi10}
{Srivastava}, A.~K. 2010, Advances in Geosciences, Volume 21: Solar Terrestrial
  (ST), 21, 315

\bibitem[\protect\citeauthoryear{{Srivastava} et~al.}{{Srivastava}
  et~al.}{2008}]{Sri08}
{Srivastava}, A.~K., {Zaqarashvili}, T.~V., {Uddin}, W., {Dwivedi}, B.~N.,  \&
  {Kumar}, P. 2008, \mnras, 388, 1899

\bibitem[\protect\citeauthoryear{{Warren} et~al.}{{Warren}
  et~al.}{1999}]{War99}
{Warren}, H.~P., {Bookbinder}, J.~A., {Forbes}, T.~G., {Golub}, L., {Hudson},
  H.~S., {Reeves}, K.,  \& {Warshall}, A. 1999, \apjl, 527, L121

\end{thebibliography}

\clearpage



\begin{figure*}
\begin{center}
\mbox{
\includegraphics[scale=0.40, angle=90]{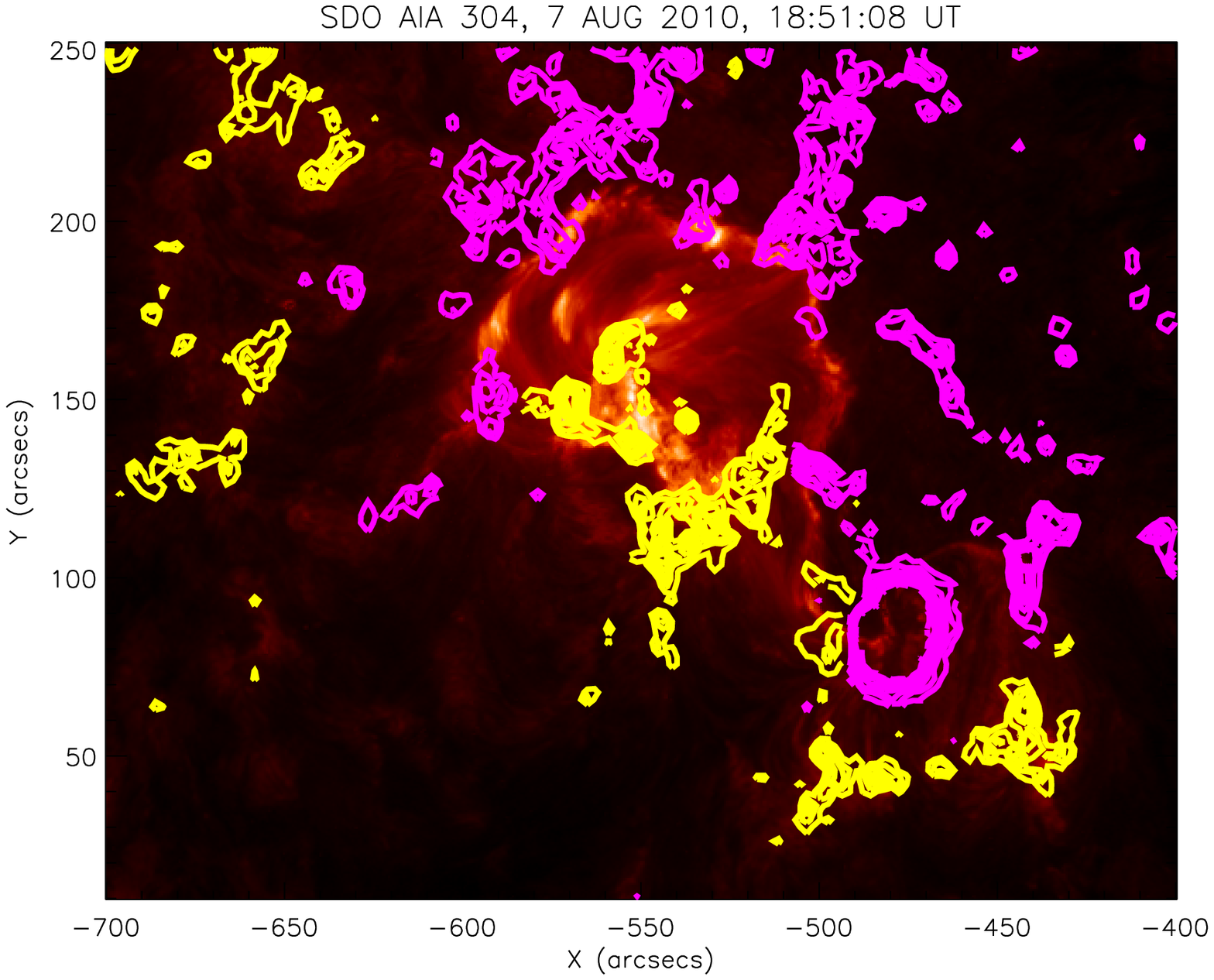}
\includegraphics[scale=0.40, angle=90]{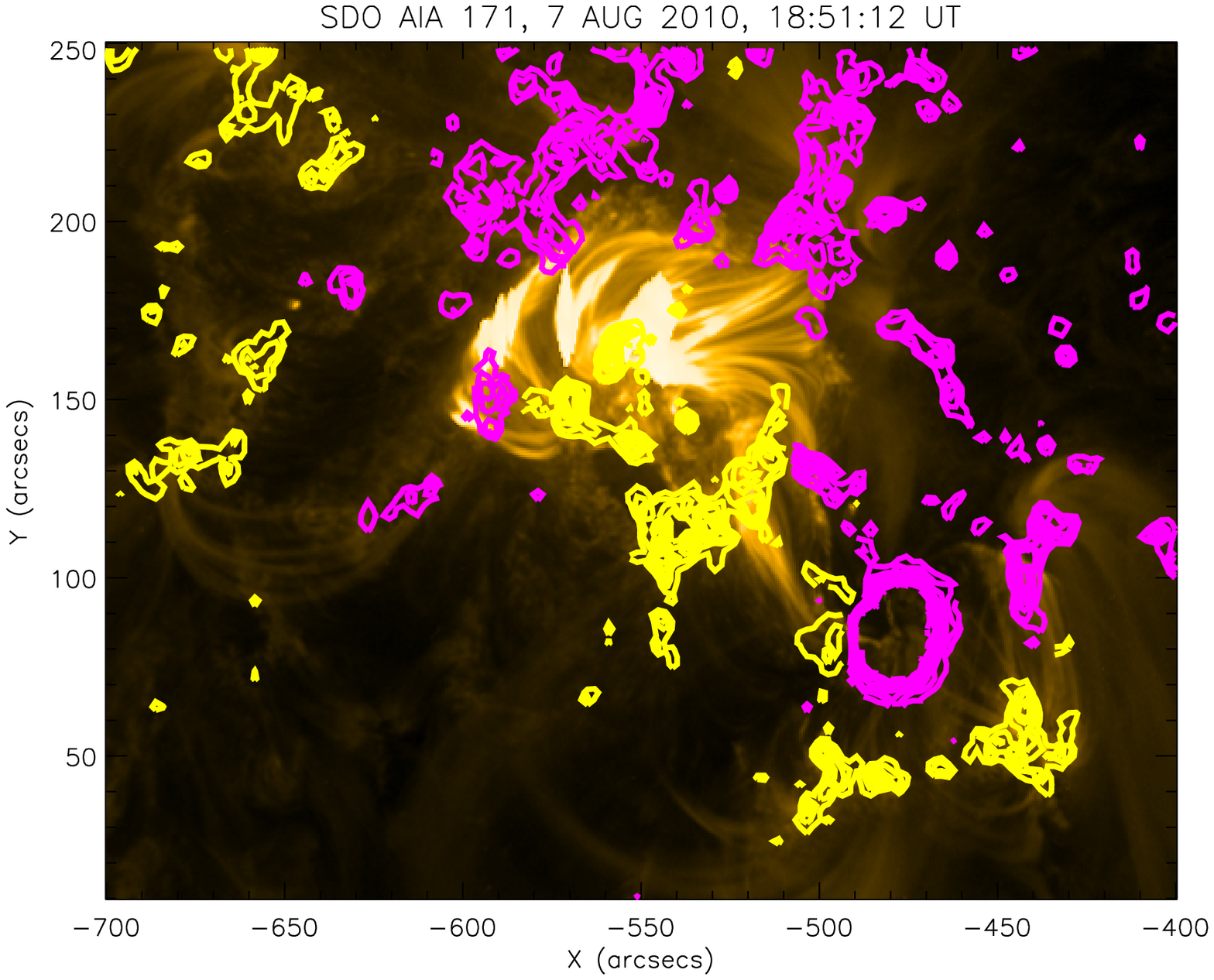}
}
\caption{\small
The enlarged field-of-view of the two-ribbons and a post flare phase of an M1.0 class 
flare in AR11093 are shown at 18:51 UT on 7 August 2010 as observed 
by SDO/AIA 304 \AA\ (left-panel) and 171 \AA\ (right-panel). The SoHO/MDI
magnetic polarities are over-plotted above these snapshots that clearly reveal
the opposite field distributions (+ve with yellow and -ve with magenta) at the
two flare ribbons. The contour levels are with $\pm{800}$ G 
magnetic field strength.}
\label{fig:initial_profile}
\end{center}
\end{figure*}

\clearpage

\begin{figure*}
\centering
\mbox{
\includegraphics[scale=0.35, angle=90]{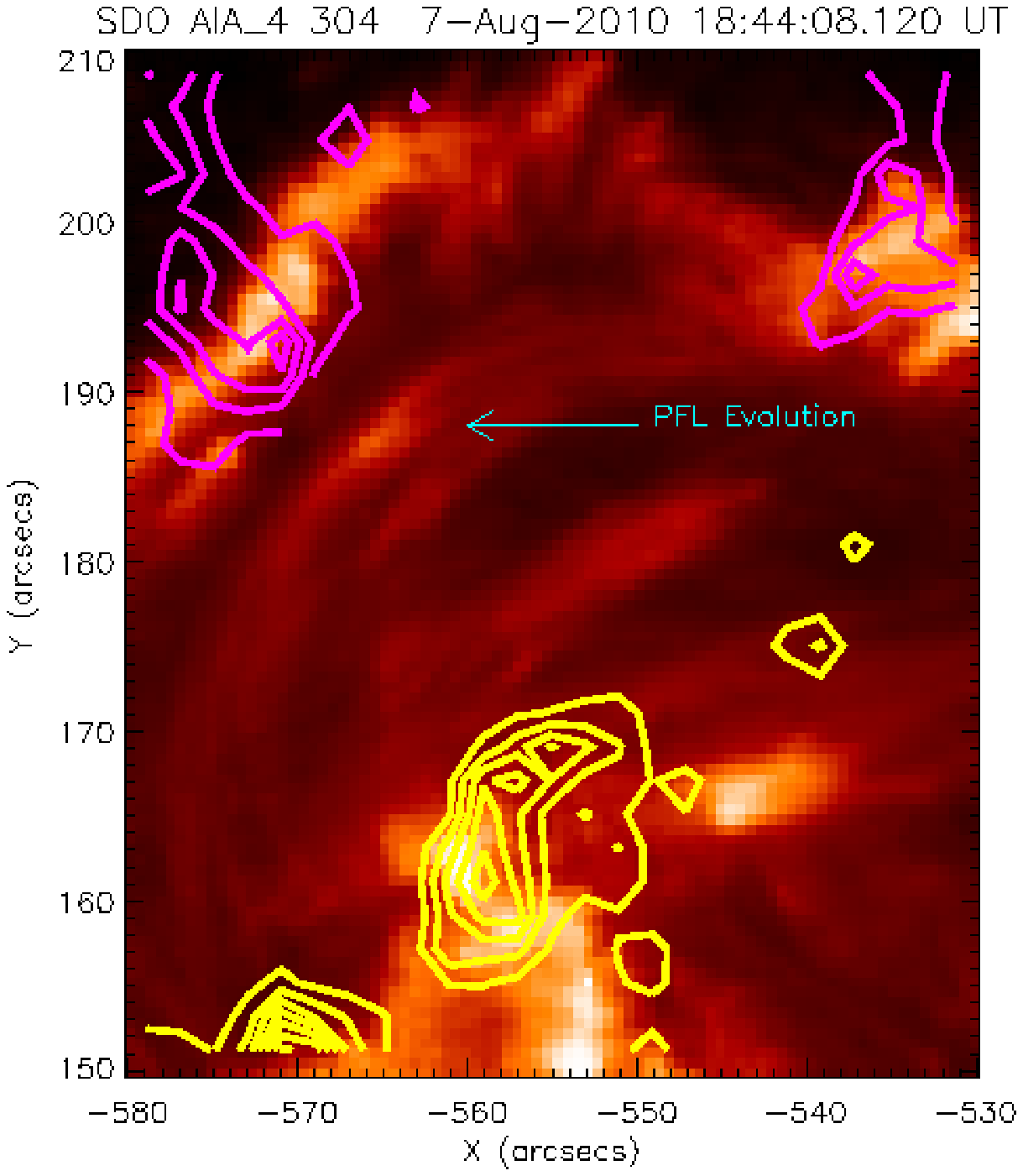}
\includegraphics[scale=0.35, angle=90]{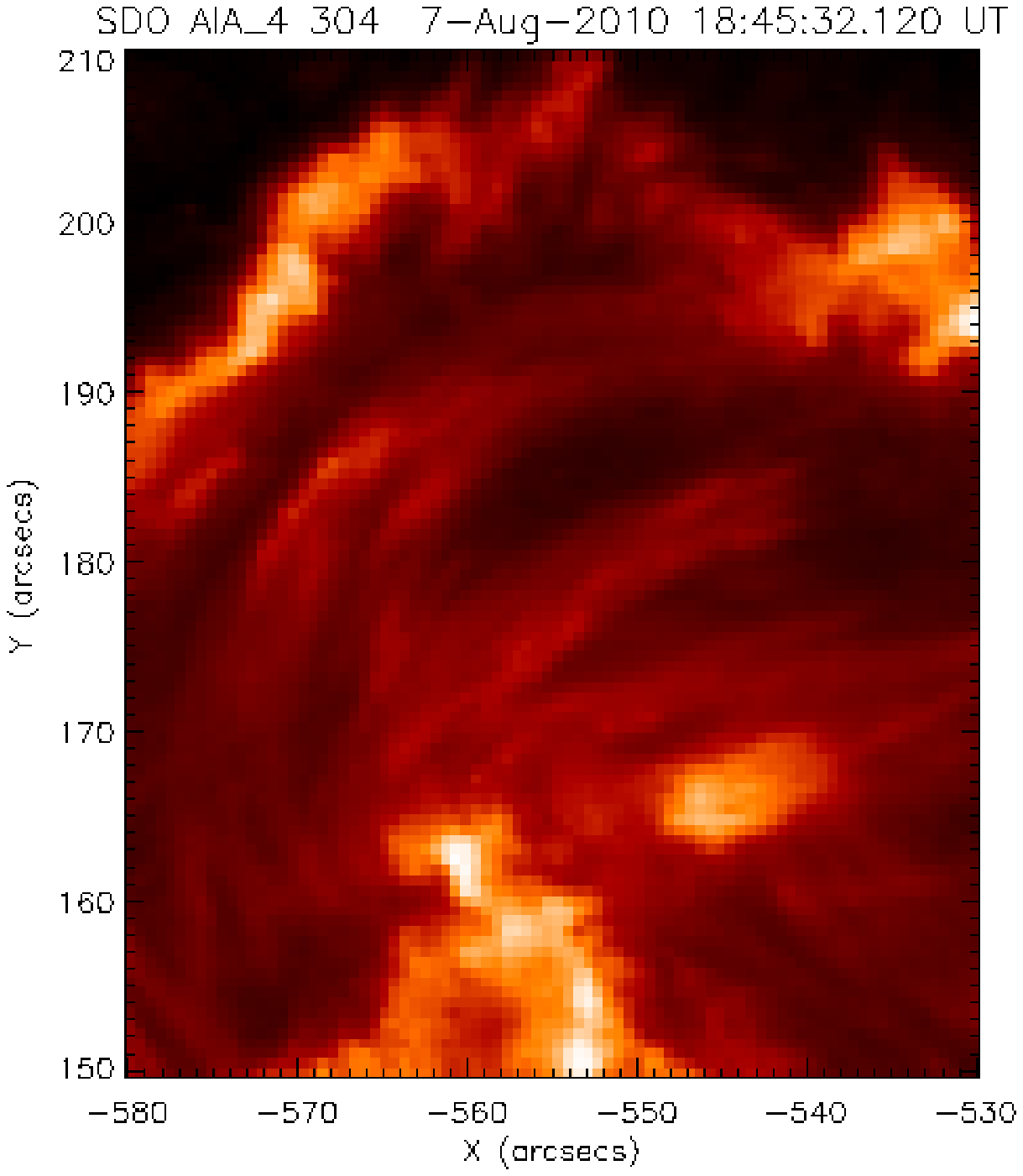}
\includegraphics[scale=0.35, angle=90]{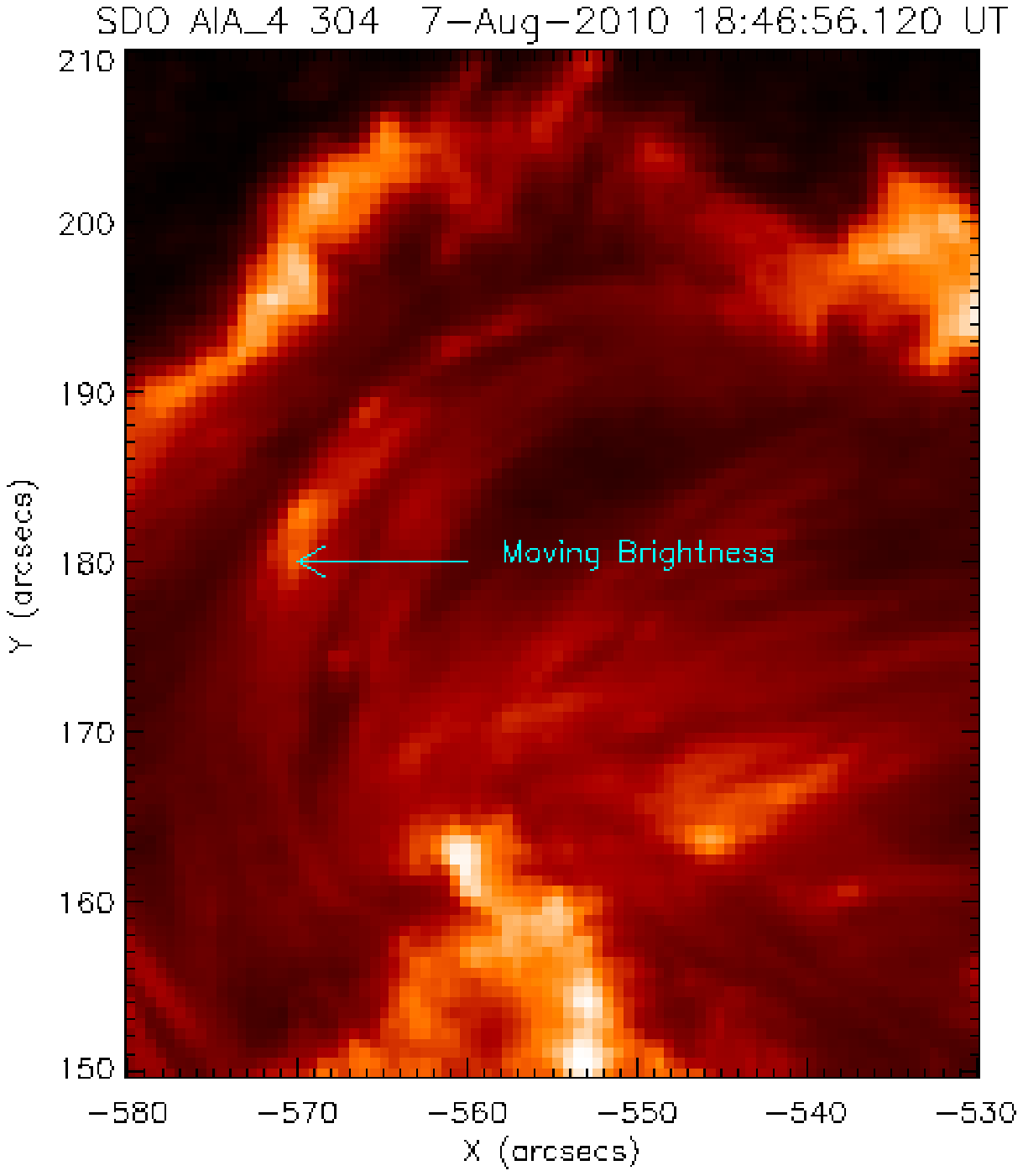}}
\mbox{
\includegraphics[scale=0.35, angle=90]{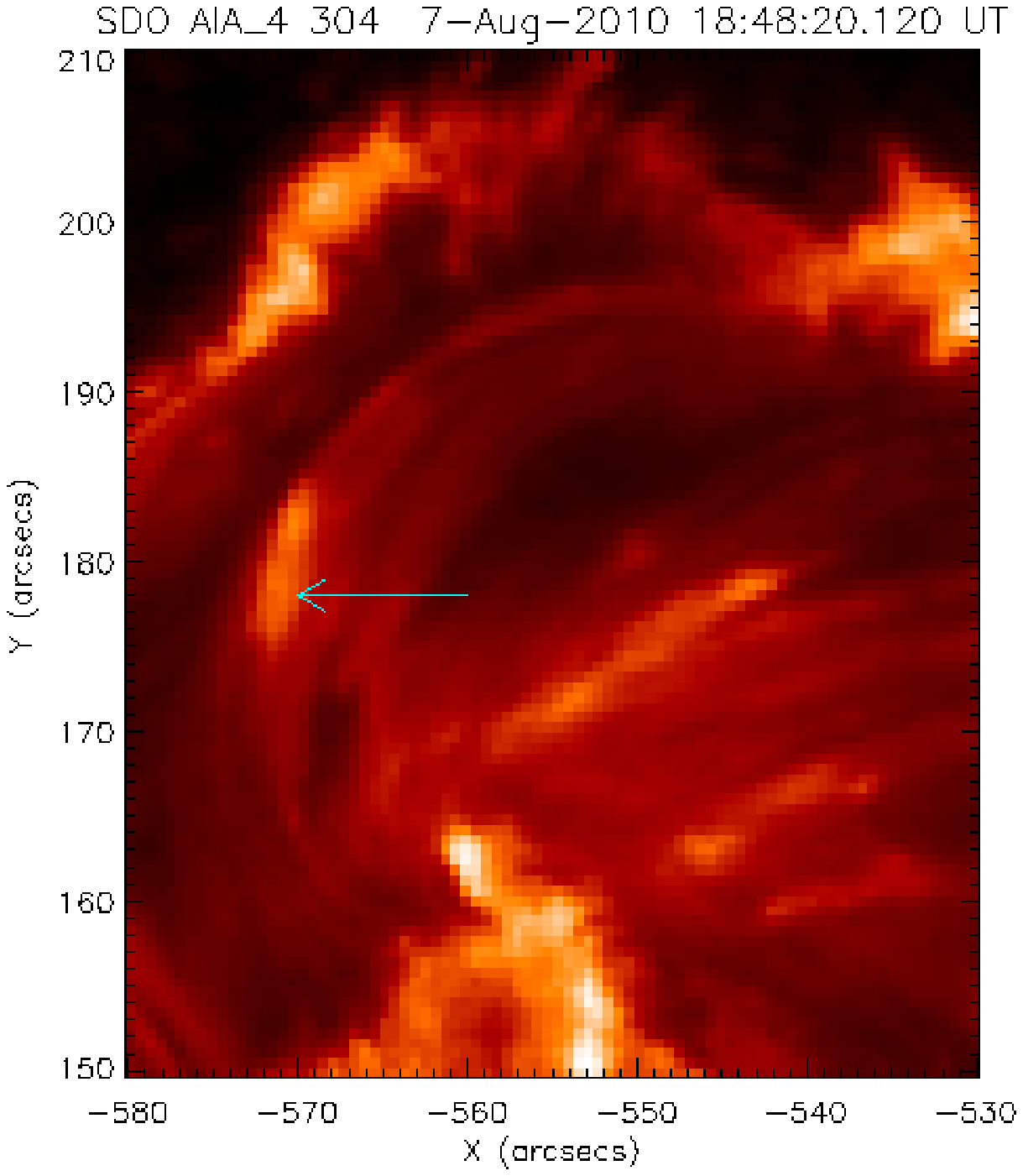}
\includegraphics[scale=0.35, angle=90]{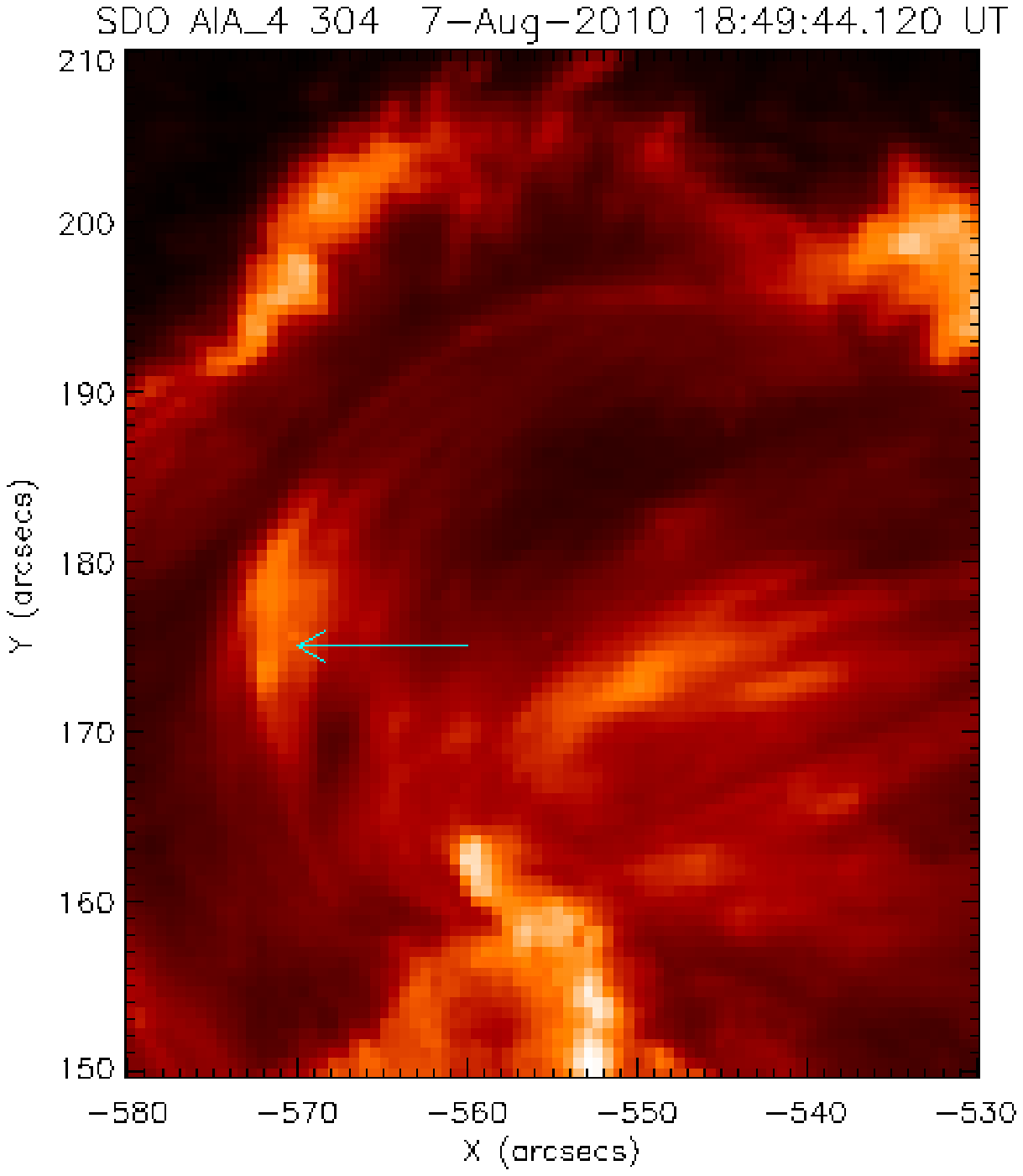}
\includegraphics[scale=0.35, angle=90]{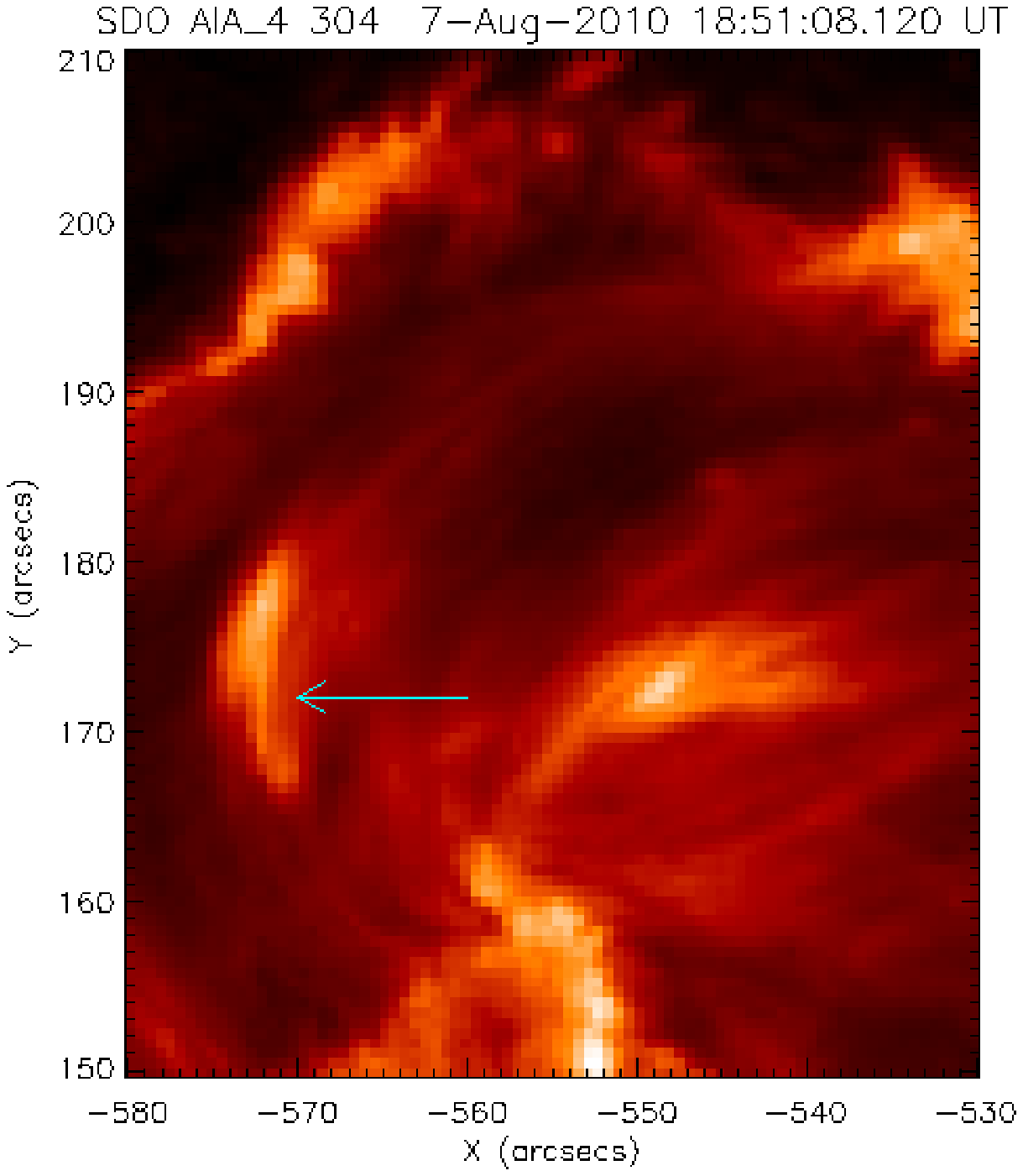}}
\mbox{
\includegraphics[scale=0.35, angle=90]{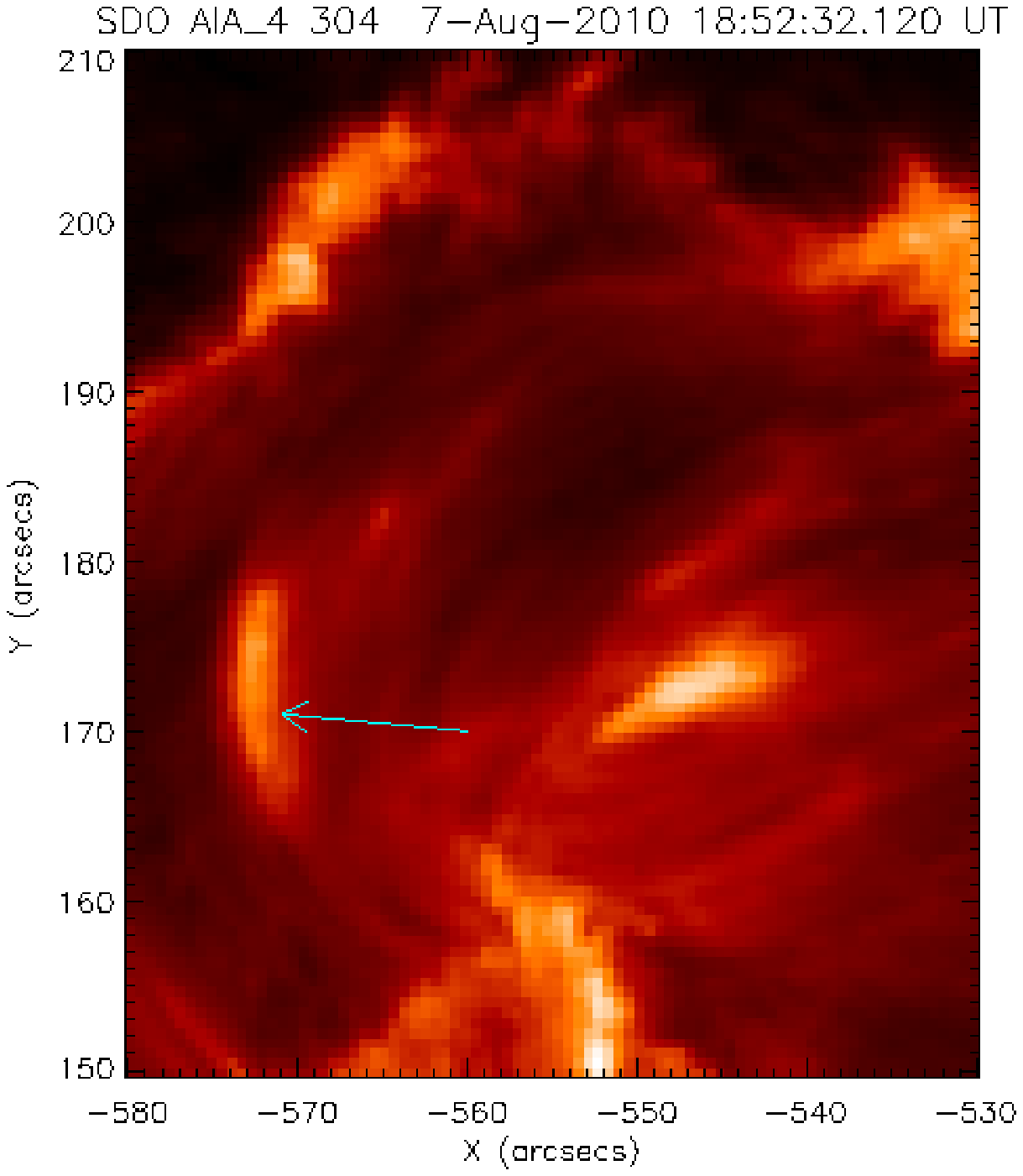}
\includegraphics[scale=0.35, angle=90]{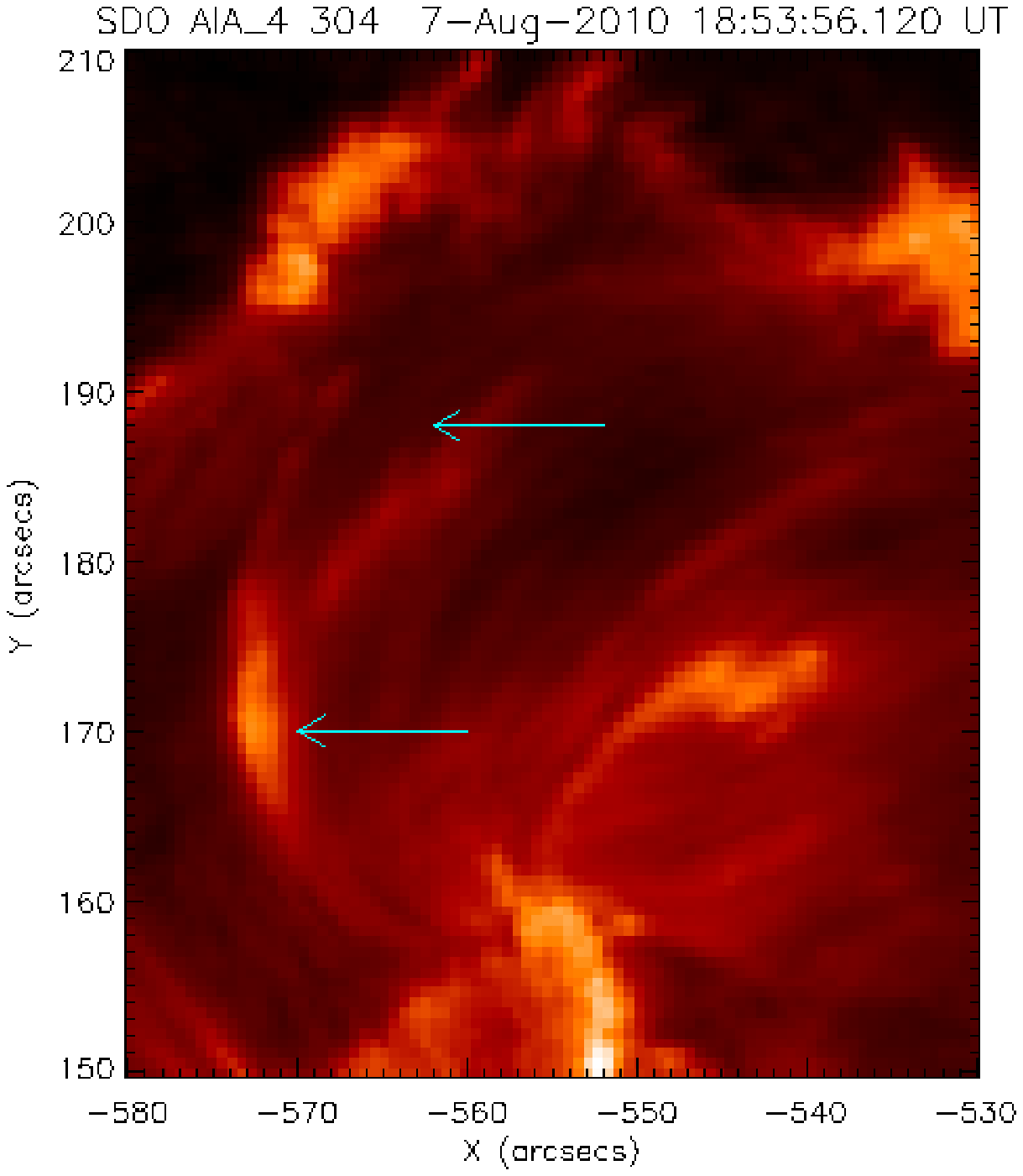}
\includegraphics[scale=0.35, angle=90]{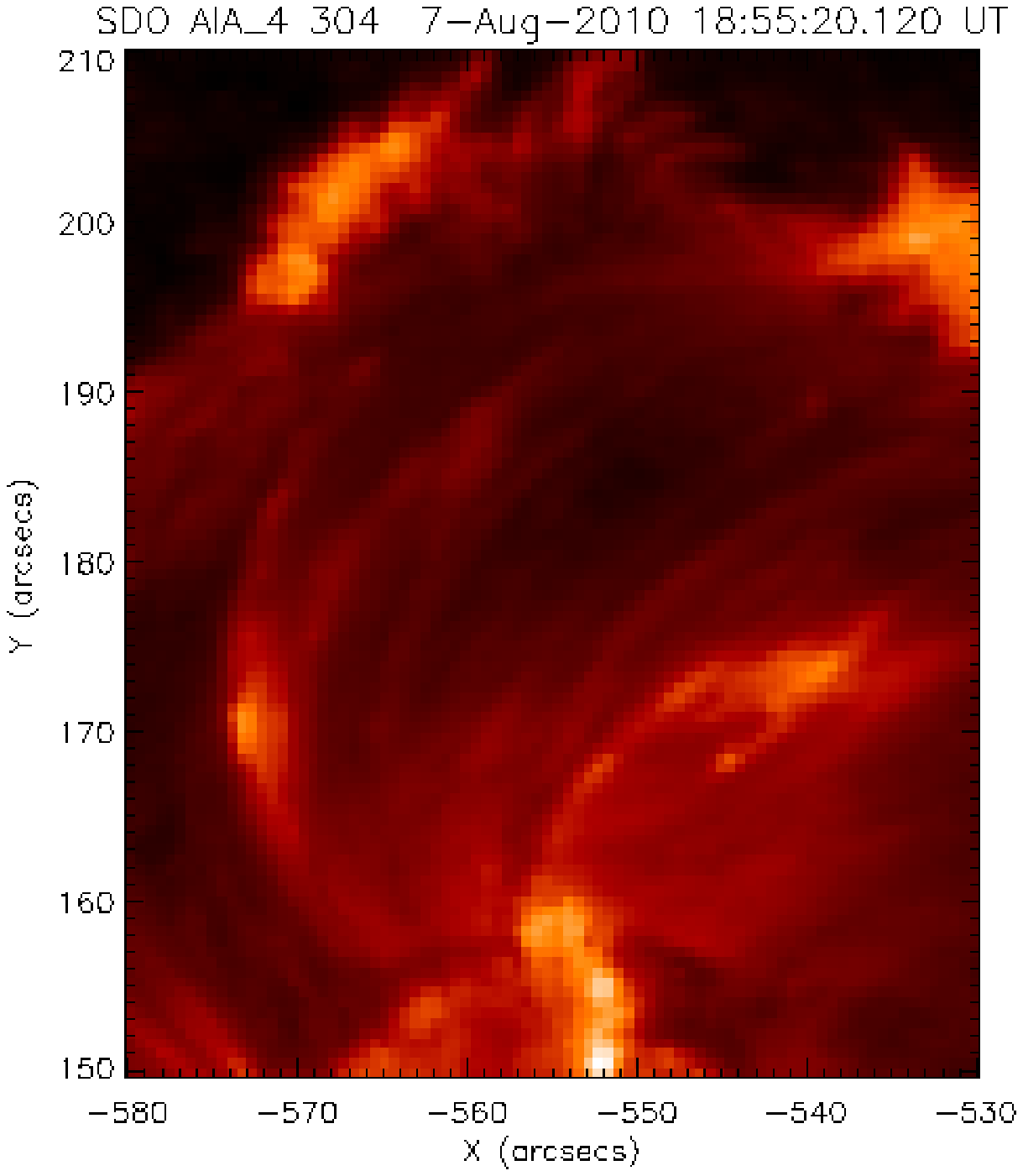}}
\mbox{
\includegraphics[scale=0.35, angle=90]{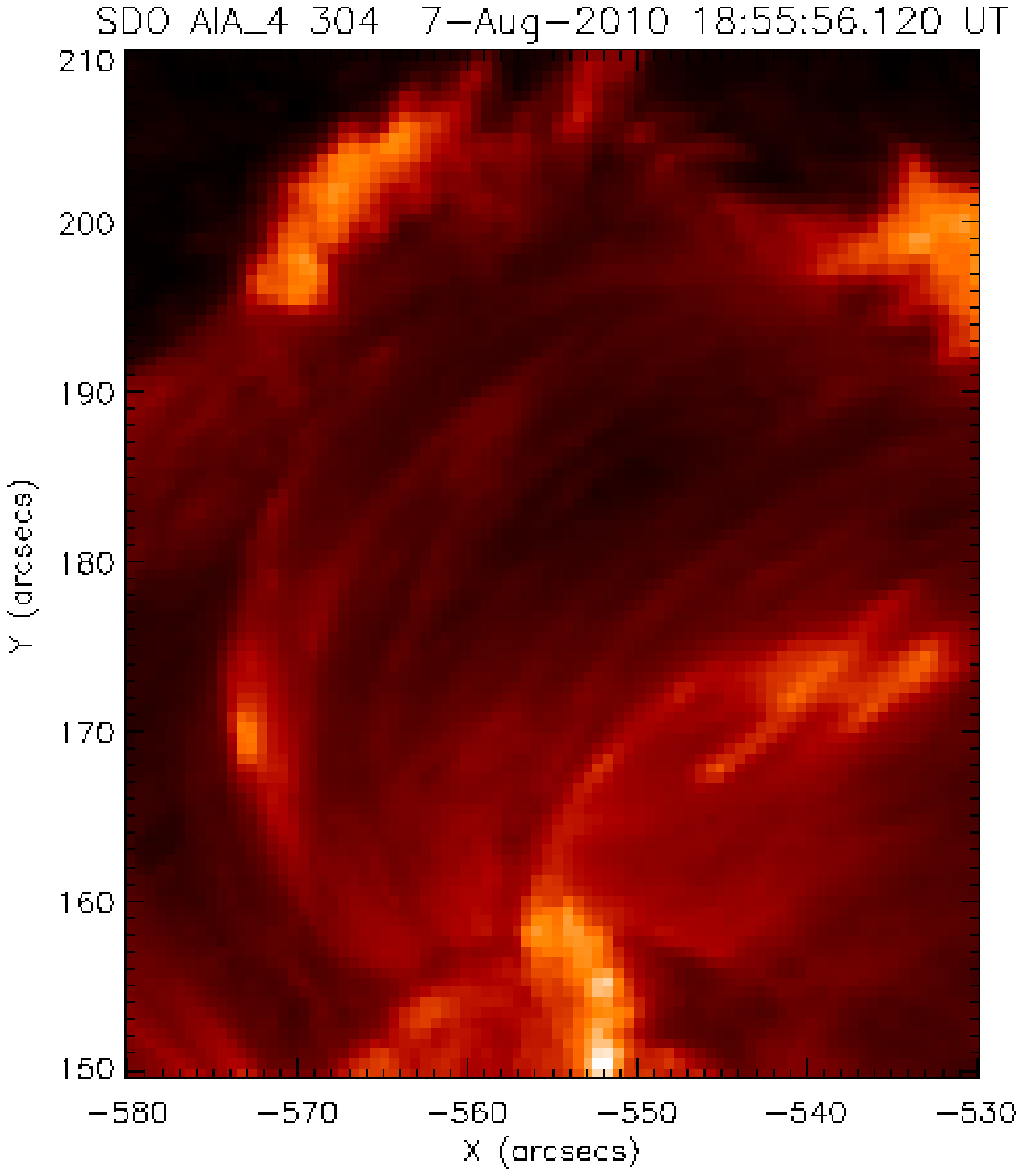}
\includegraphics[scale=0.35, angle=90]{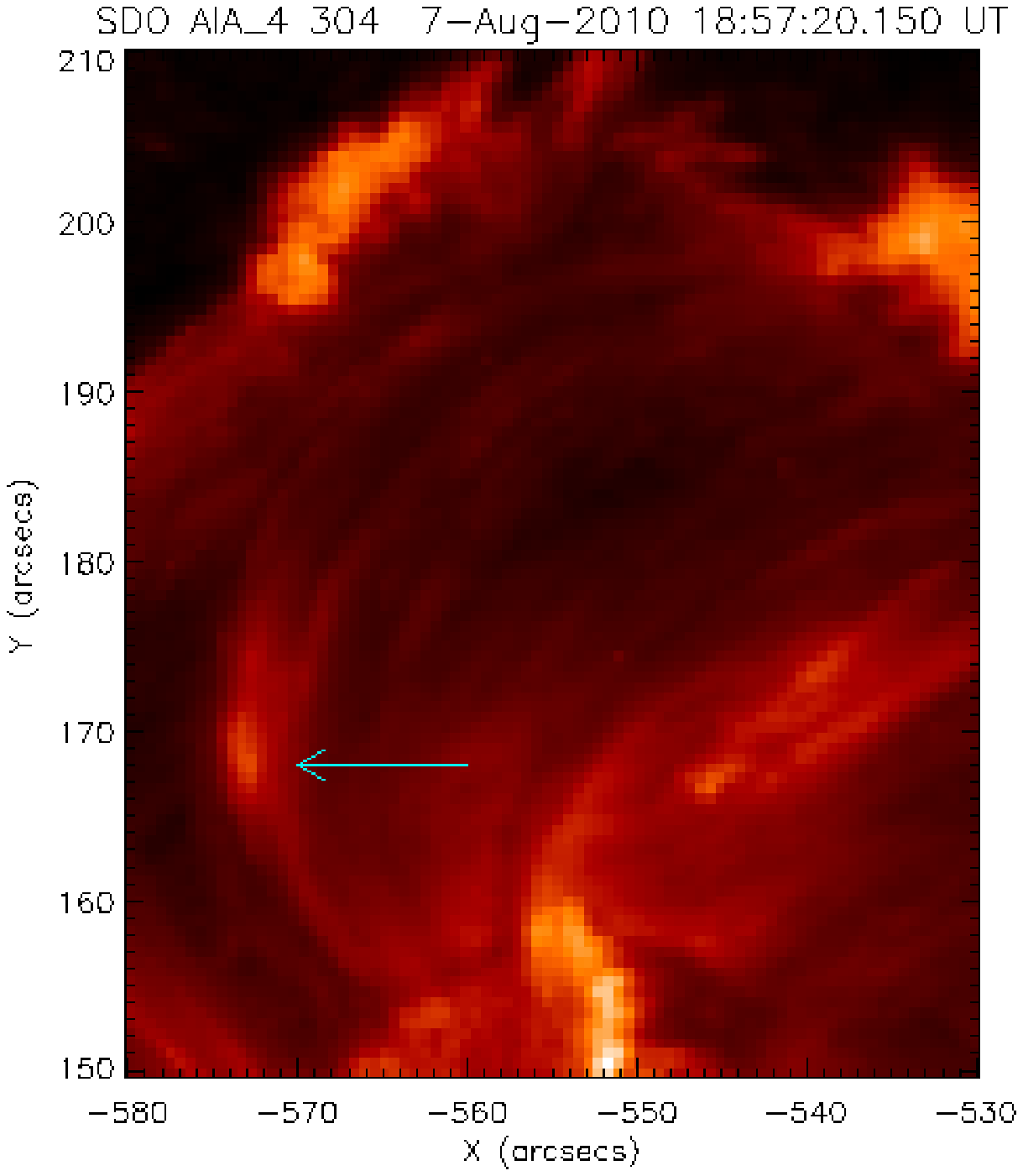}
\includegraphics[scale=0.35, angle=90]{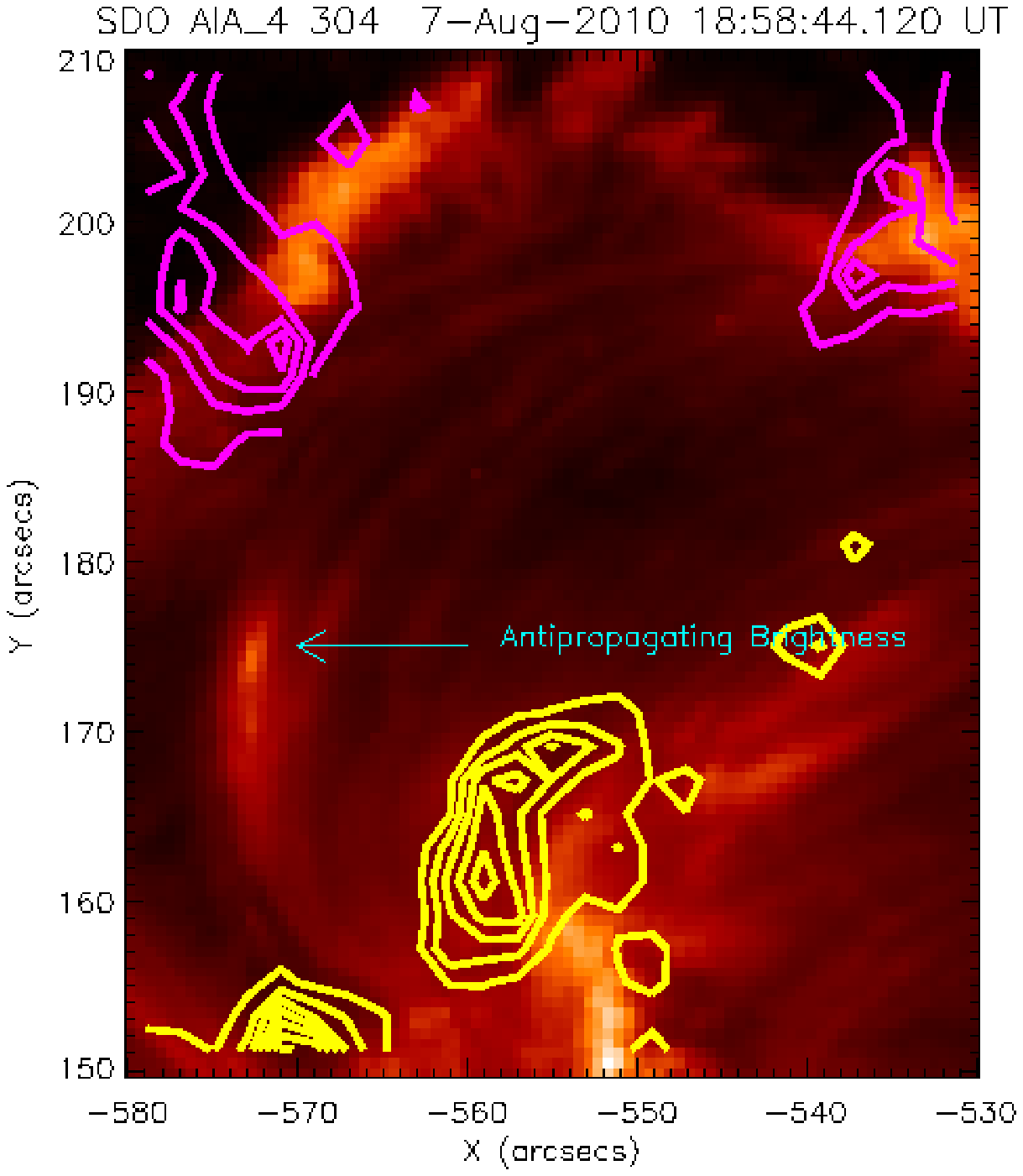}}
\caption{\small
The time sequence of a selected multi-stranded post flare loop system showing complex plasma dynamics during 18:44 UT-18:58 UT.
.The loop system and cool counter part of the bright denser plasma ($\sim$0.1 MK) are evolved on 18:44 UT. The brightness
propagates downward towards southward footpoint (18:44-18:55 UT) of the loop system and then reflects again in the northward direction
(18:58 UT) as an antipropagating brightness.}
\label{fig:JET-PULSE}
\end{figure*}

\clearpage
\begin{figure*}
\centering
\mbox{
\includegraphics[scale=0.35, angle=90]{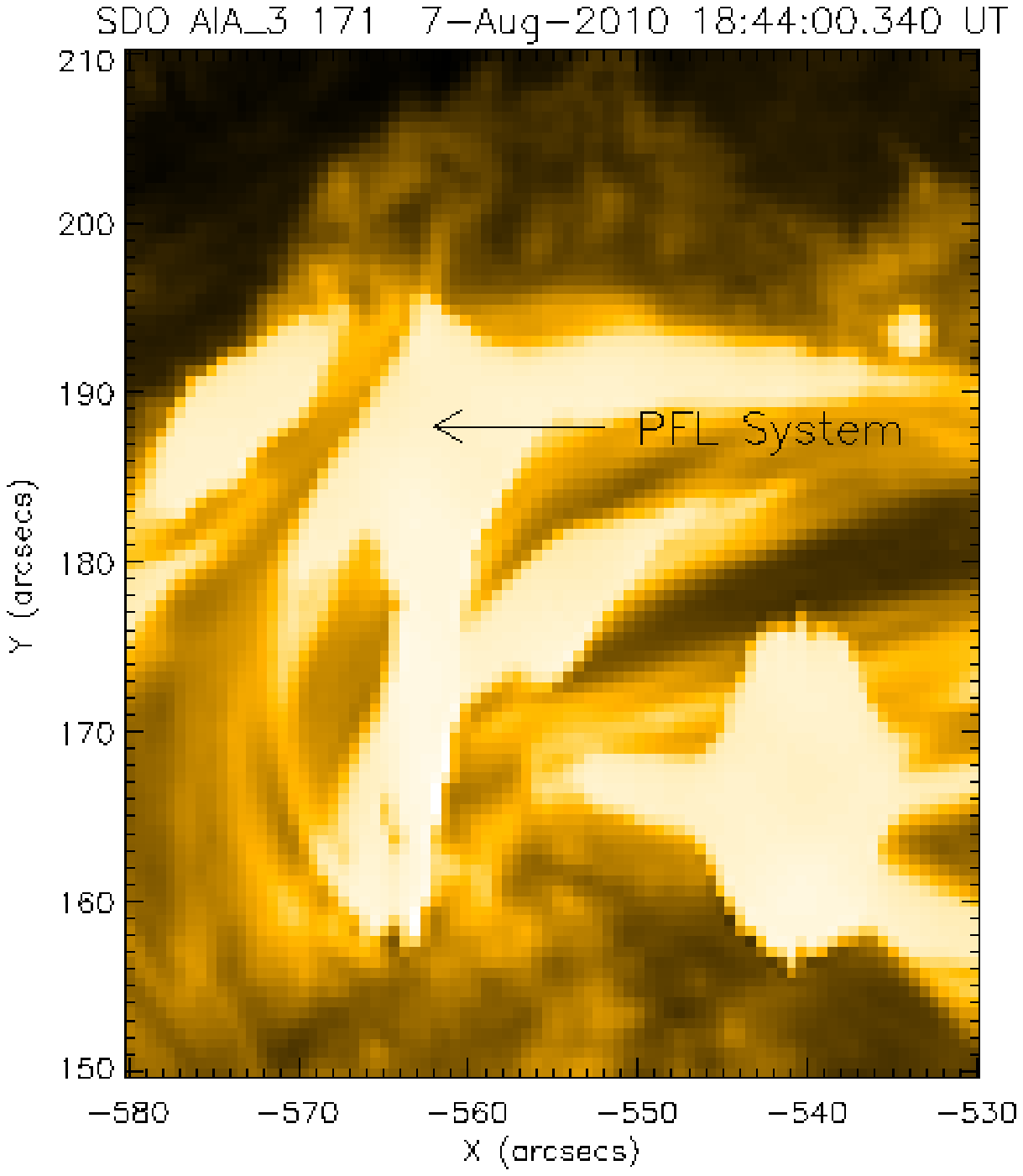}
\includegraphics[scale=0.35, angle=90]{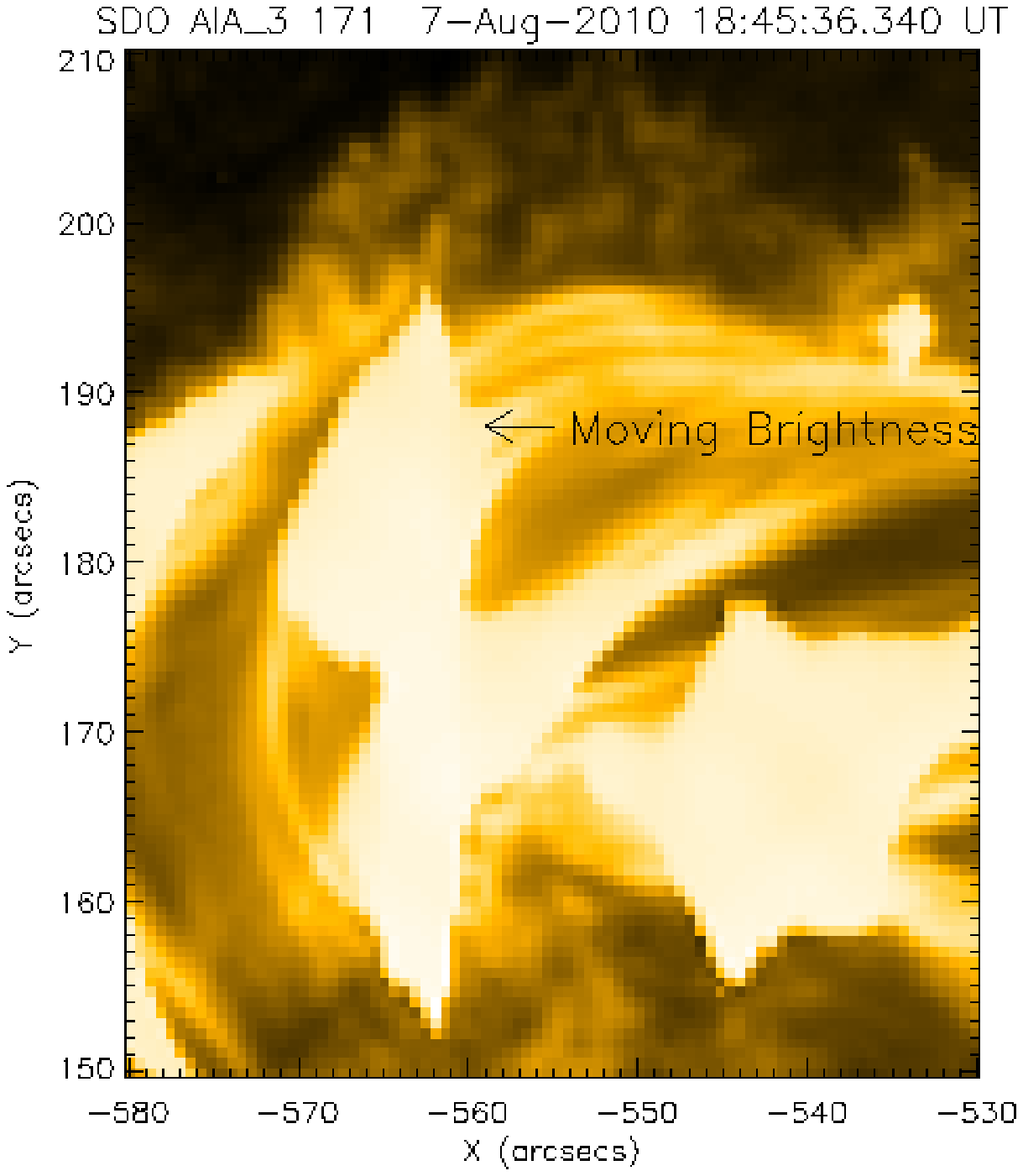}
\includegraphics[scale=0.35, angle=90]{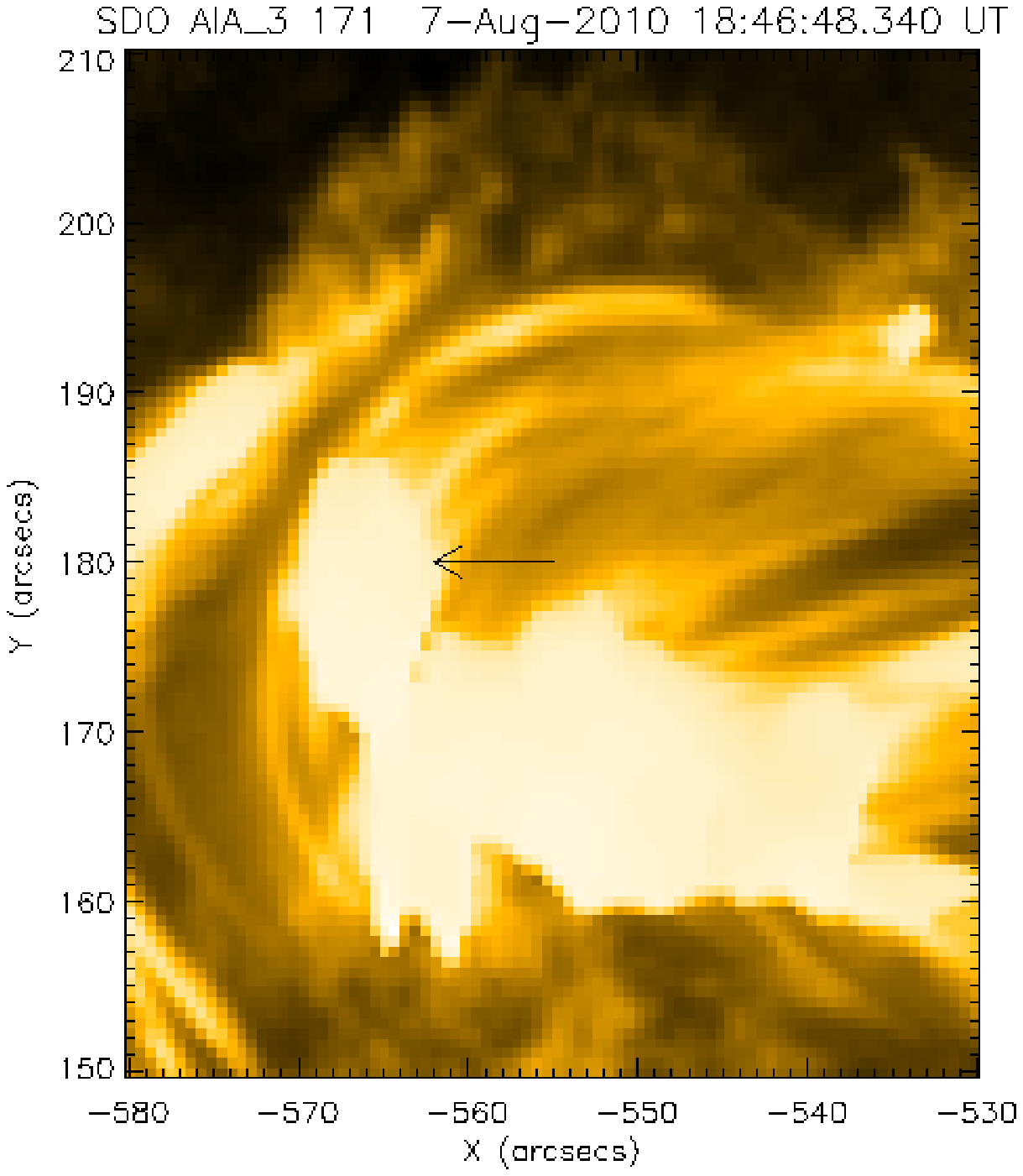}}
\mbox{
\includegraphics[scale=0.35, angle=90]{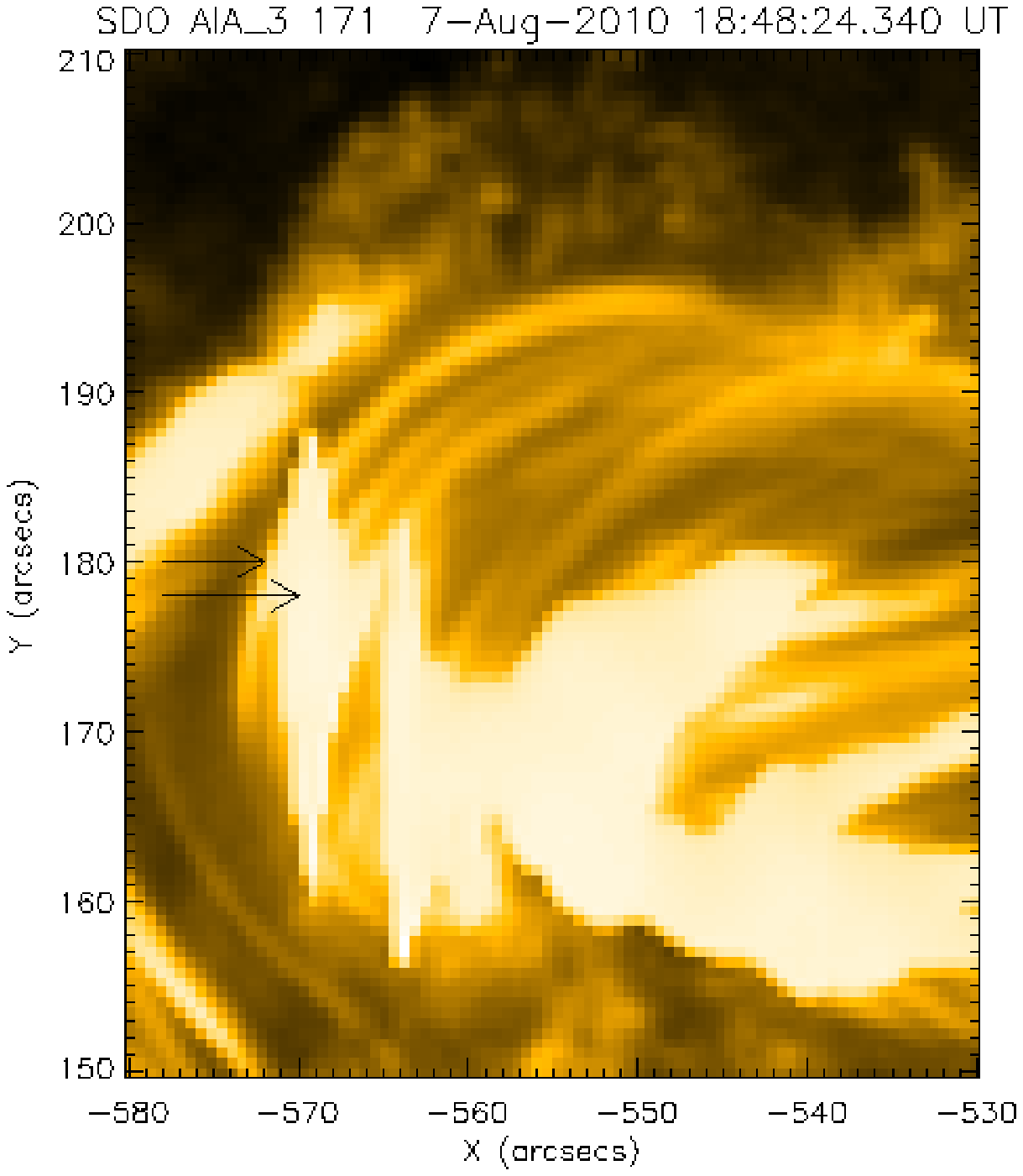}
\includegraphics[scale=0.35, angle=90]{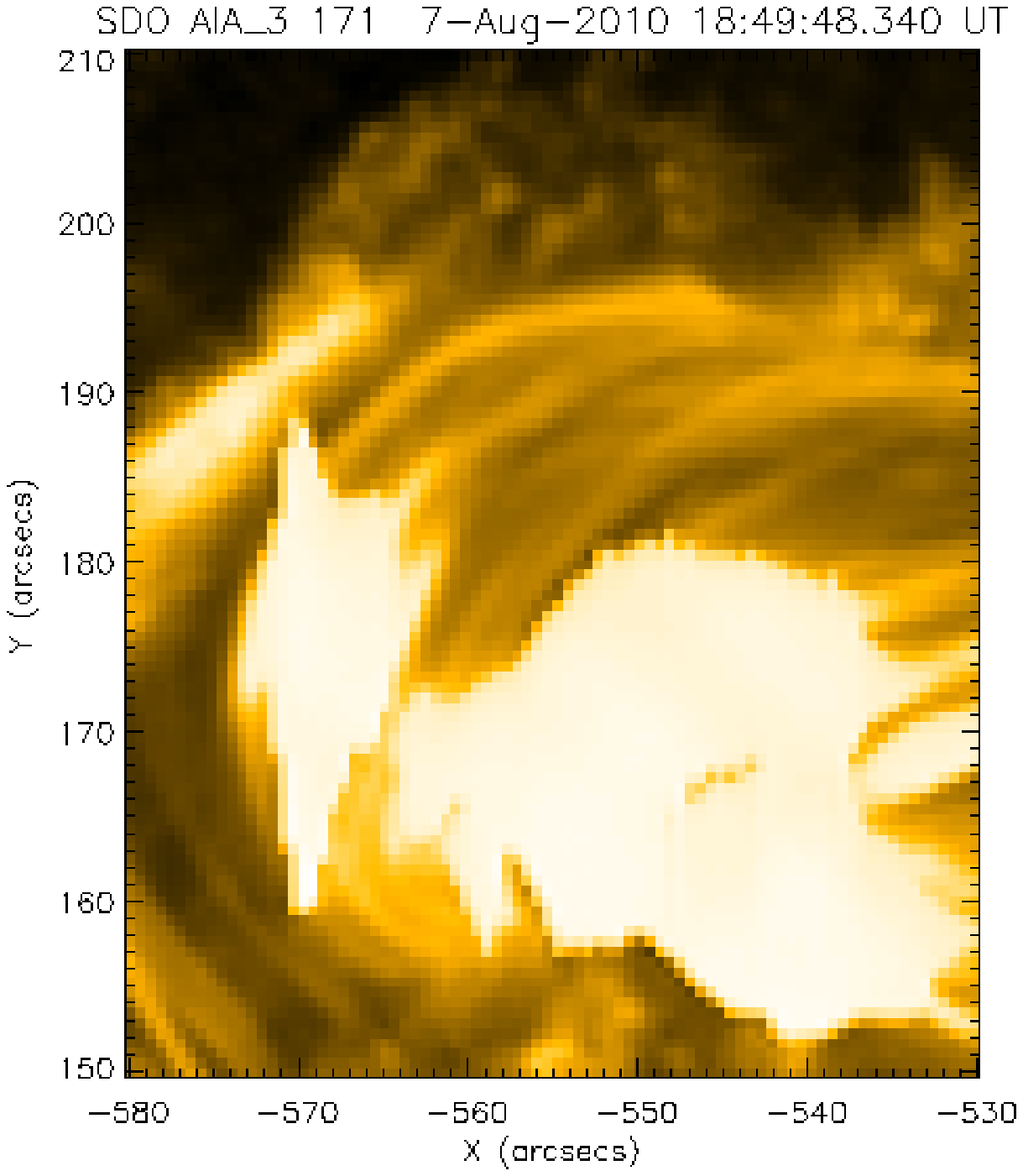}
\includegraphics[scale=0.35, angle=90]{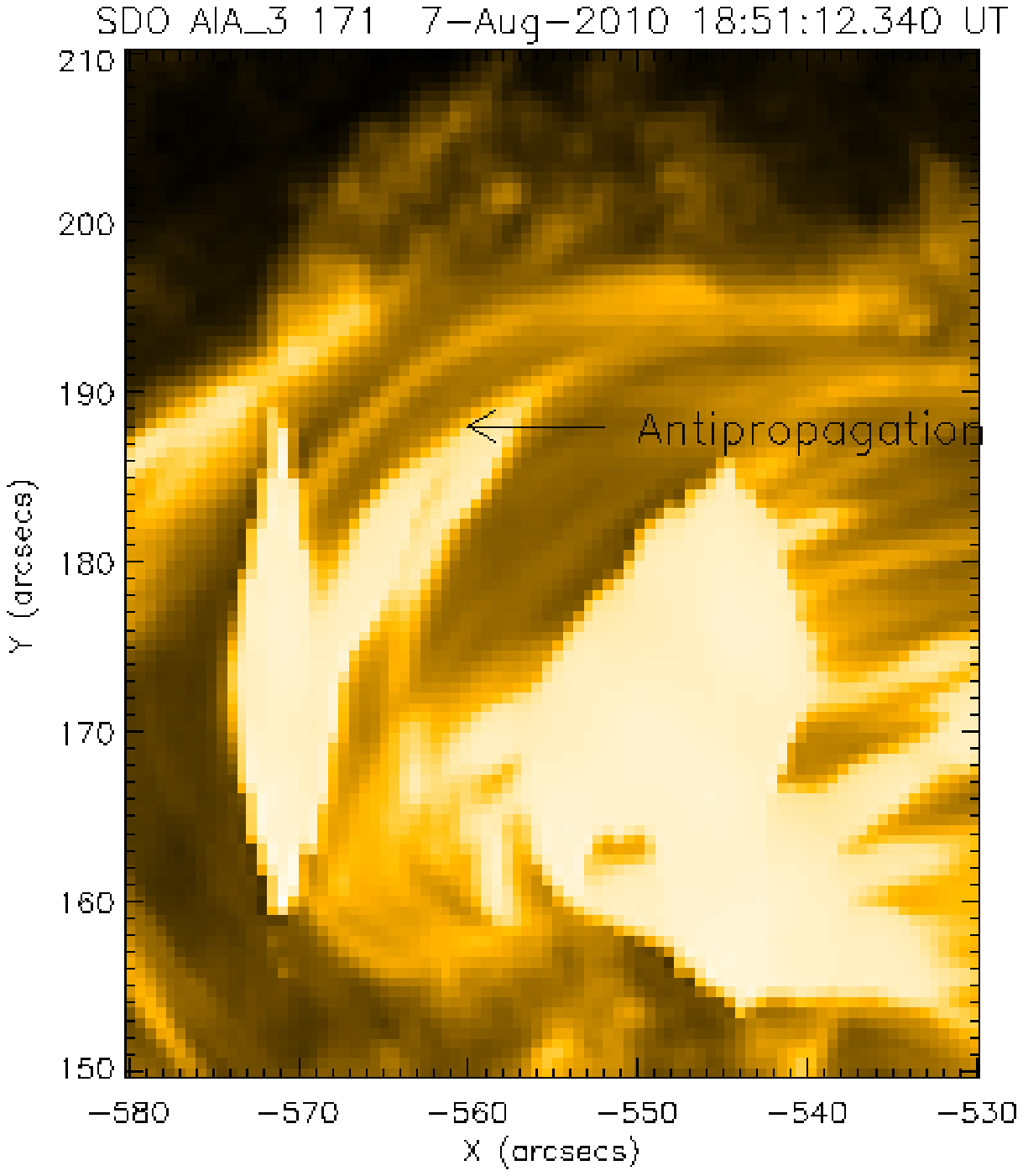}}
\mbox{
\includegraphics[scale=0.35, angle=90]{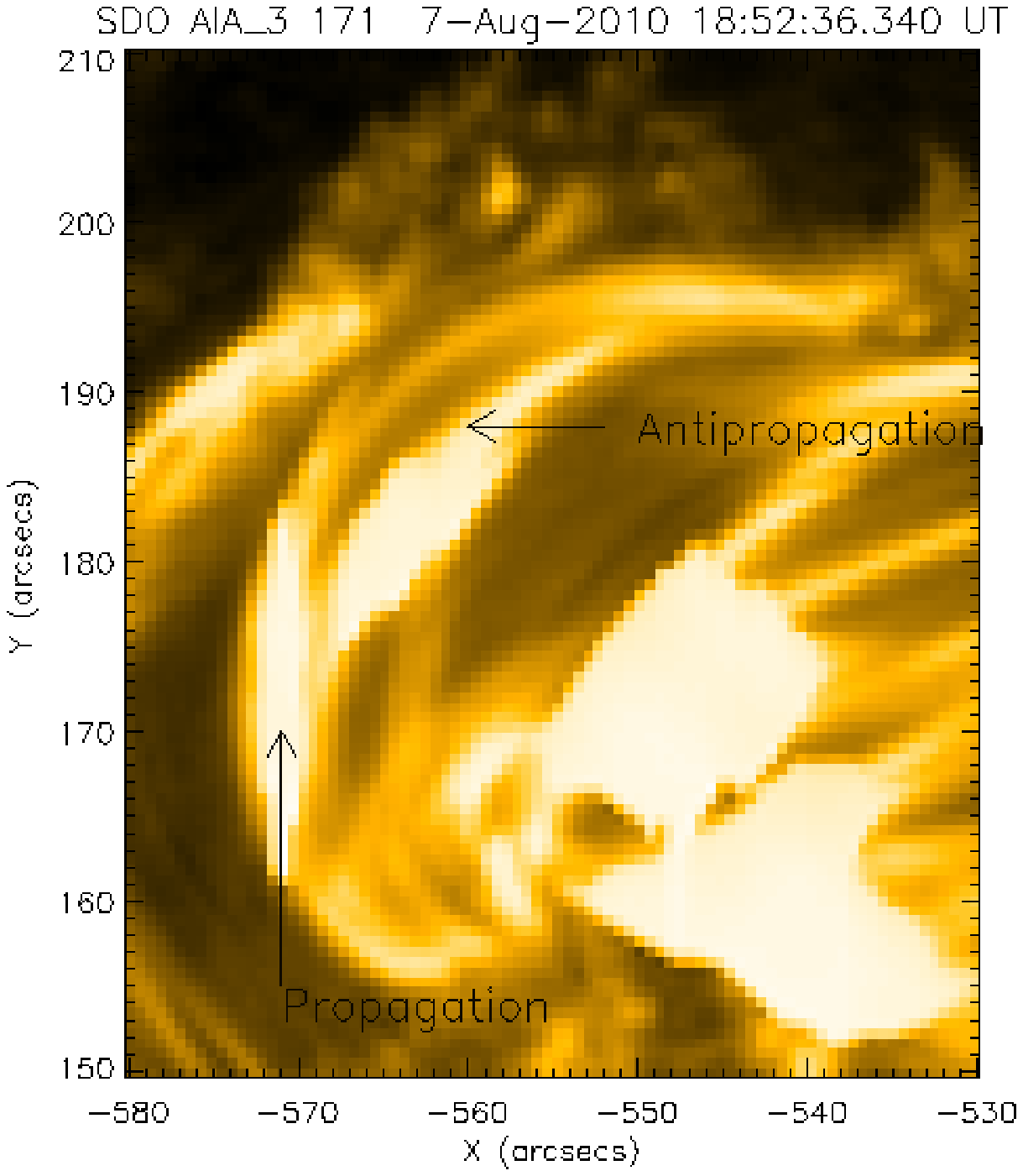}
\includegraphics[scale=0.35, angle=90]{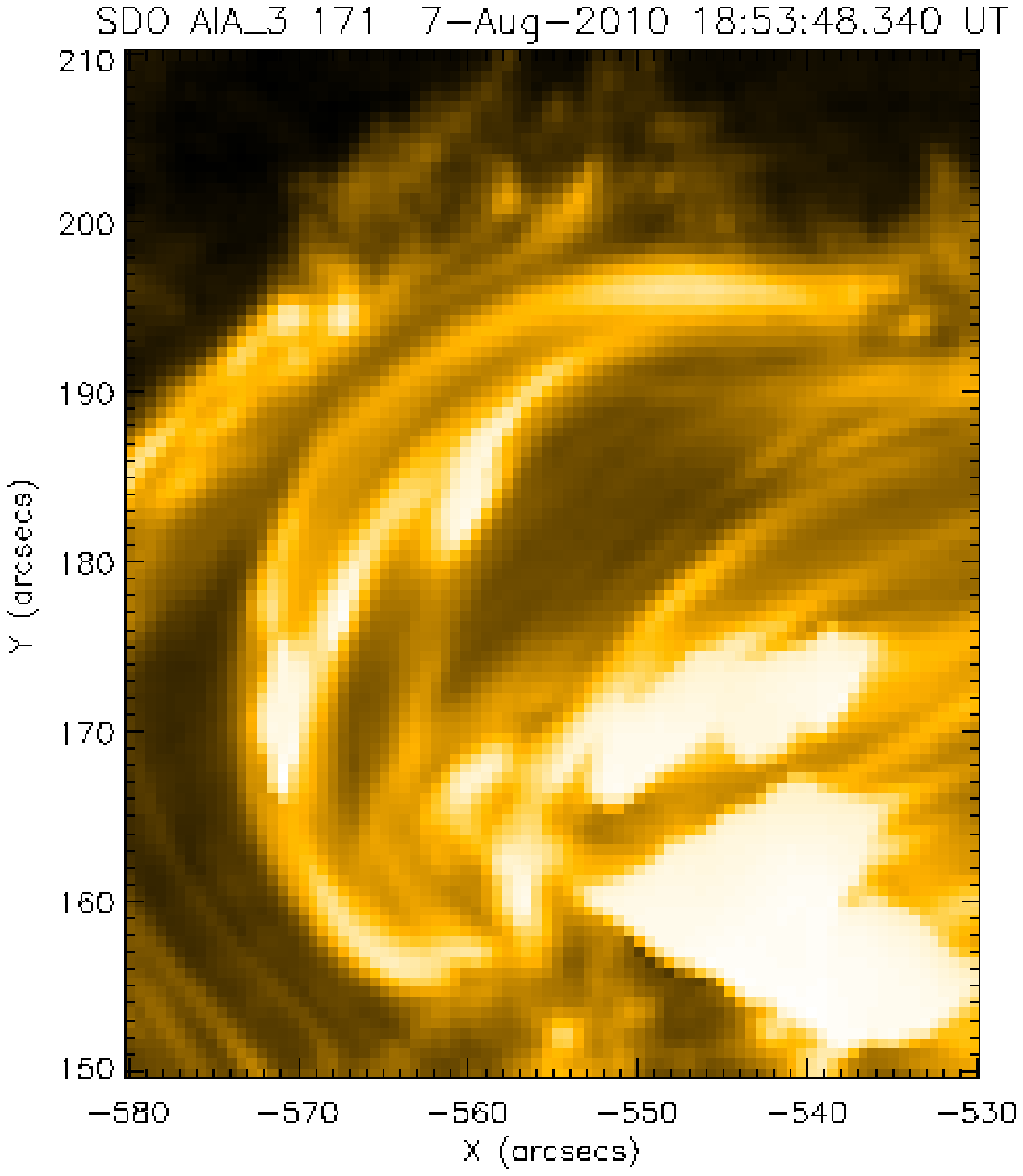}
\includegraphics[scale=0.35, angle=90]{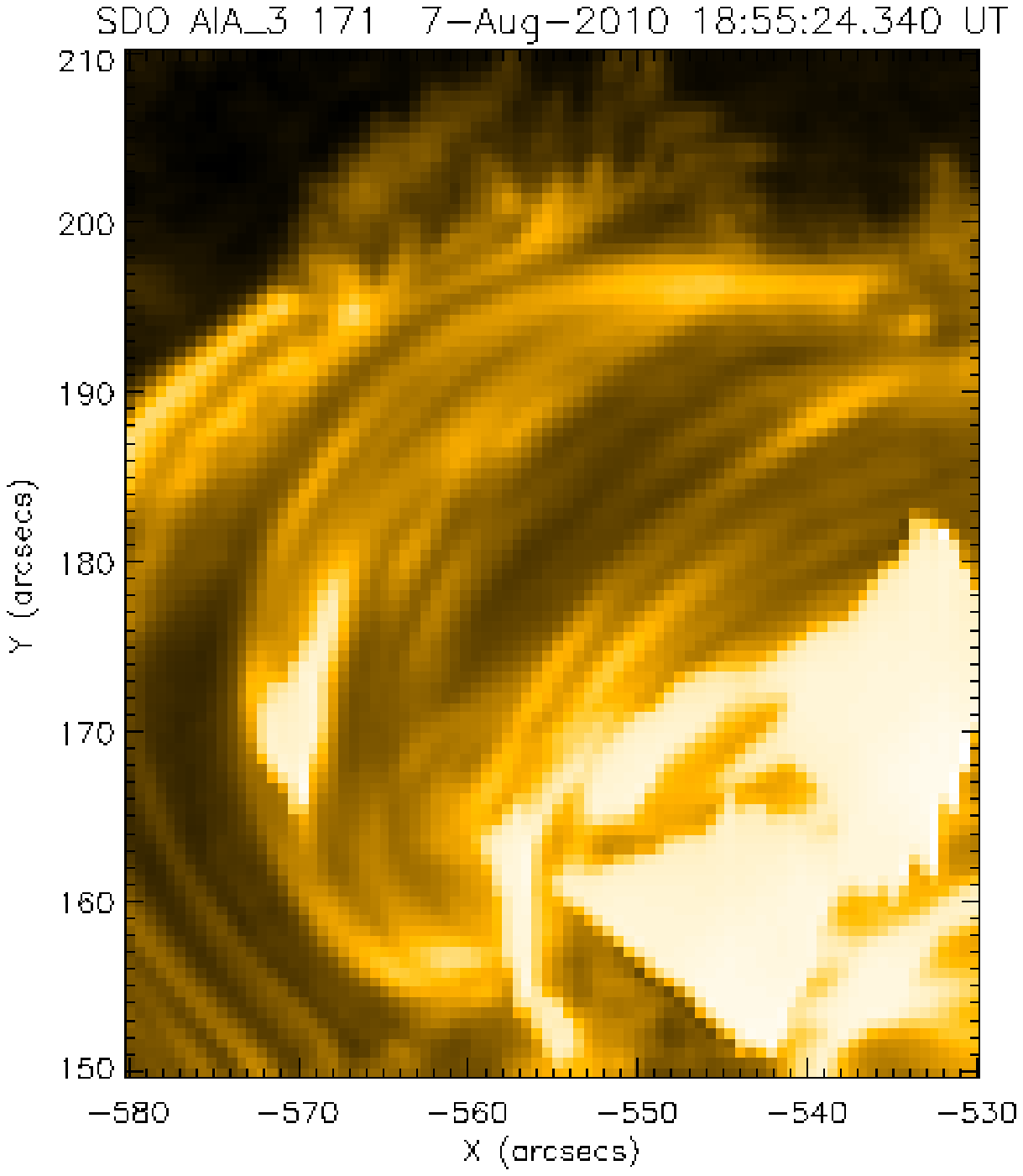}}
\mbox{
\includegraphics[scale=0.35, angle=90]{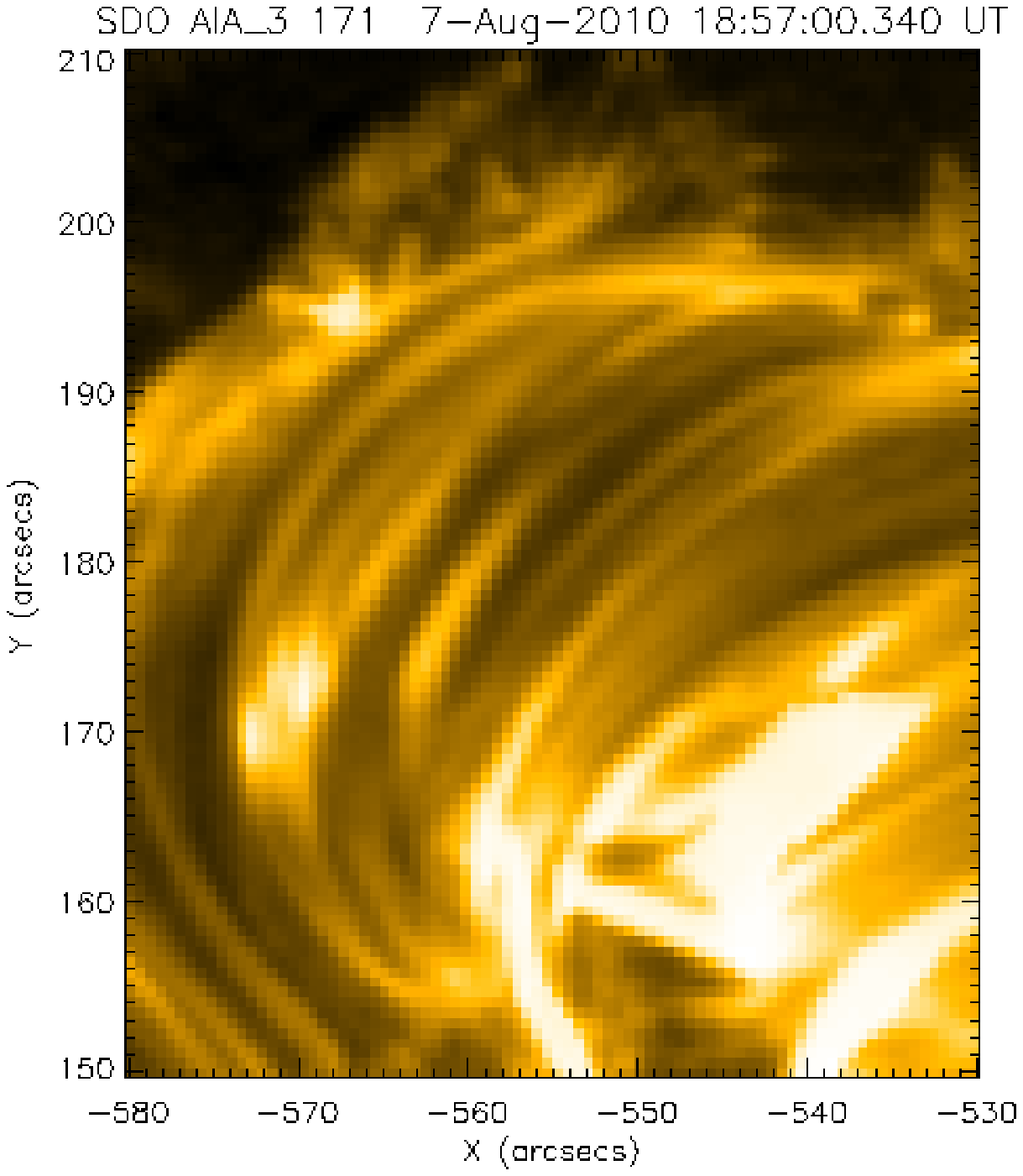}
\includegraphics[scale=0.35, angle=90]{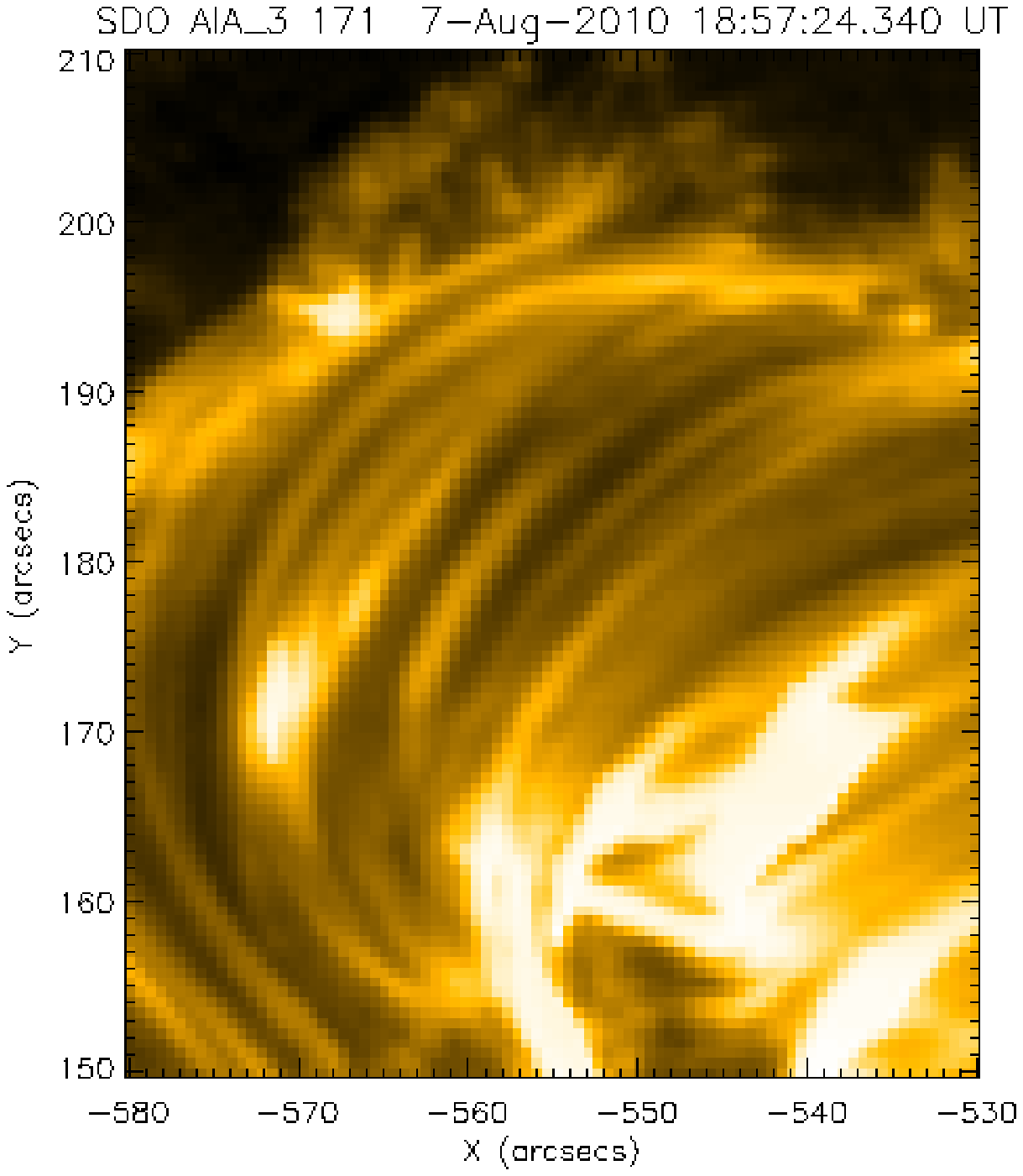}
\includegraphics[scale=0.35, angle=90]{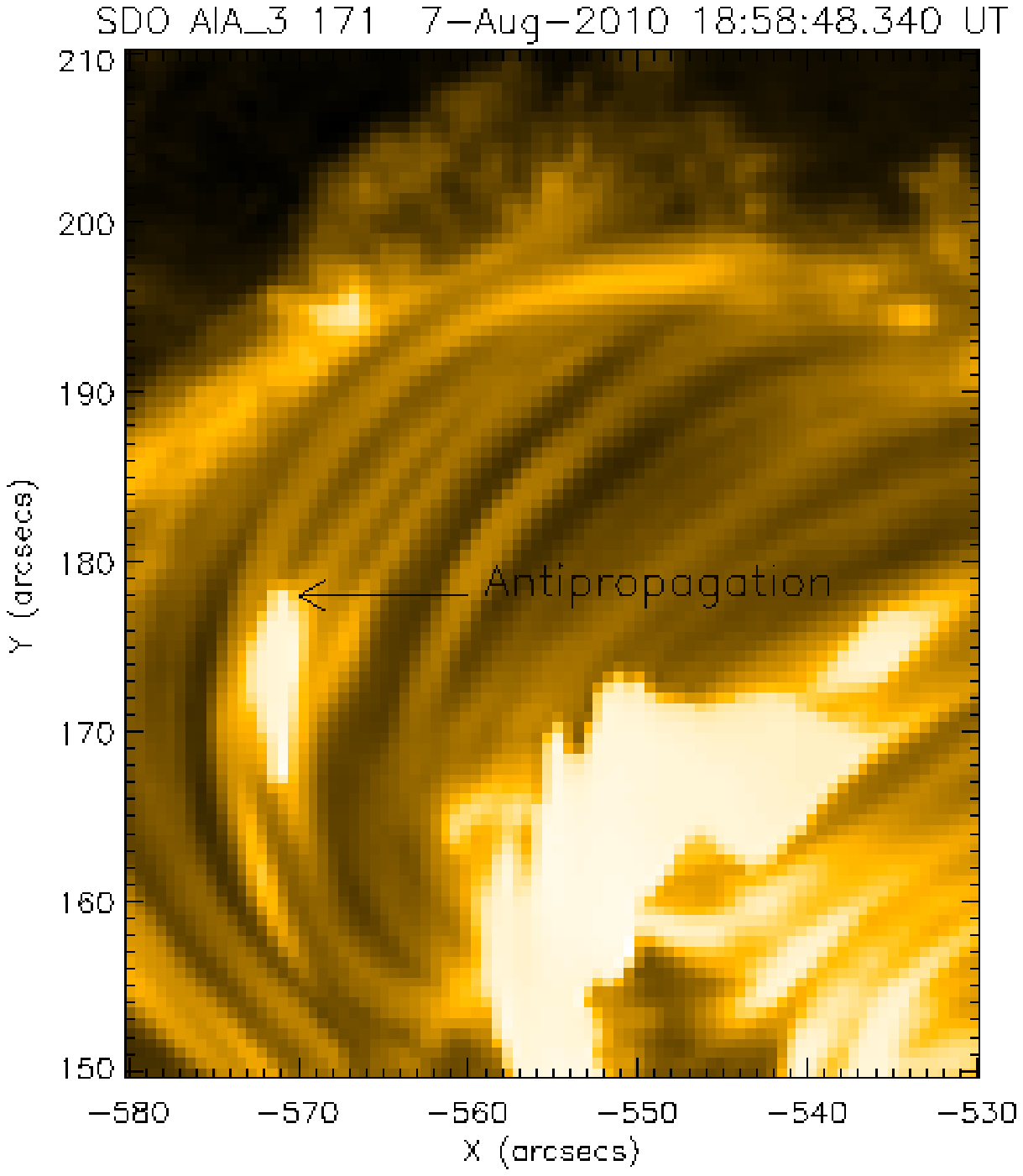}}
\caption{\small
The time sequence of a selected multi-stranded post flare loop system showing complex plasma dynamics during 18:44 UT-18:58 UT.
.The loop system and hot counter part of the bright denser plasma ($\sim$1.0 MK) are evolved on 18:44 UT. The brightness
propagates downward towards southward footpoint of the loop system and then reflects again in the northward direction
as an antipropagating brightness.}
\label{fig:JET-PULSE}
\end{figure*}

\clearpage

\begin{figure*}
\centering
\includegraphics[scale=0.21, angle=90]{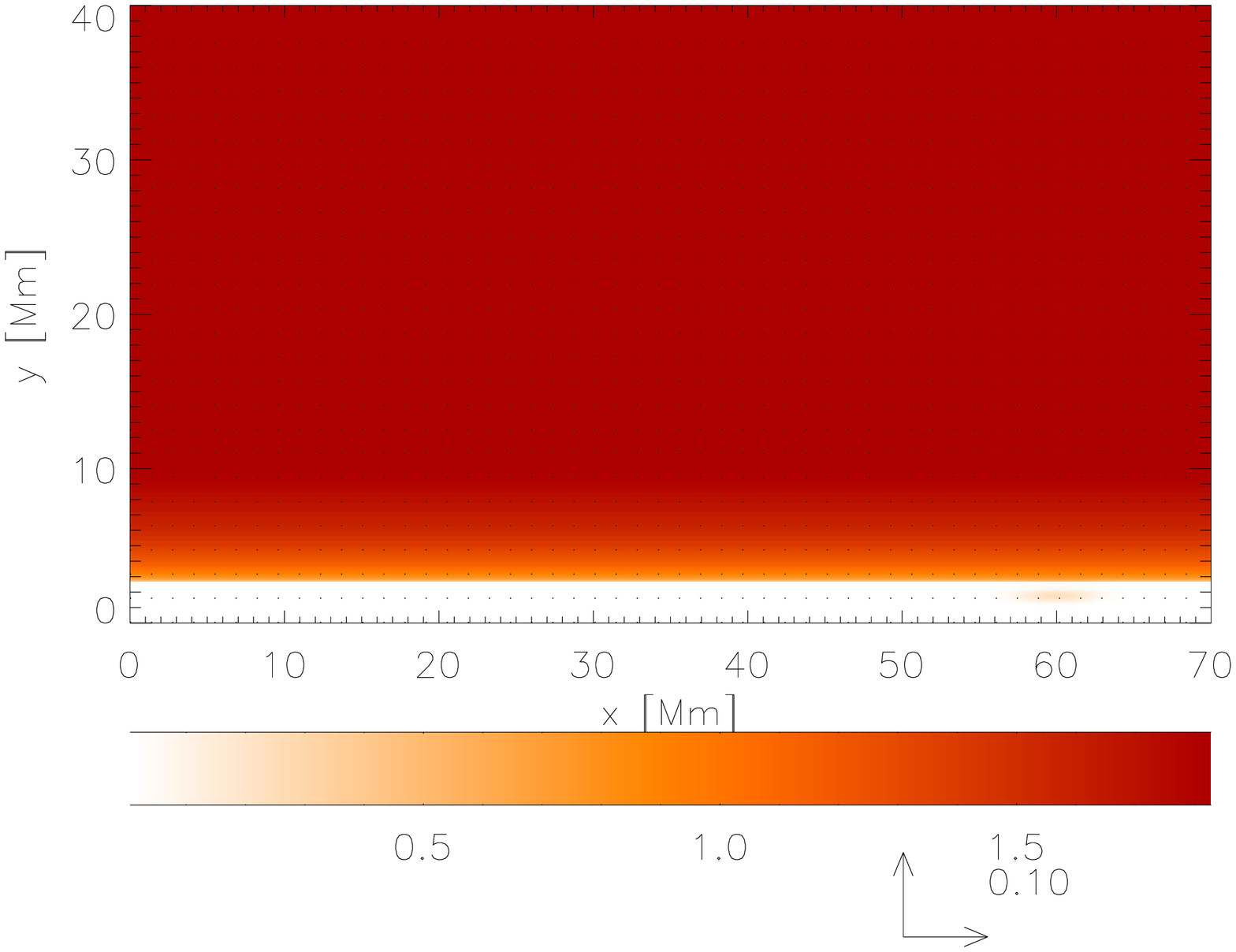}
\includegraphics[scale=0.21, angle=90]{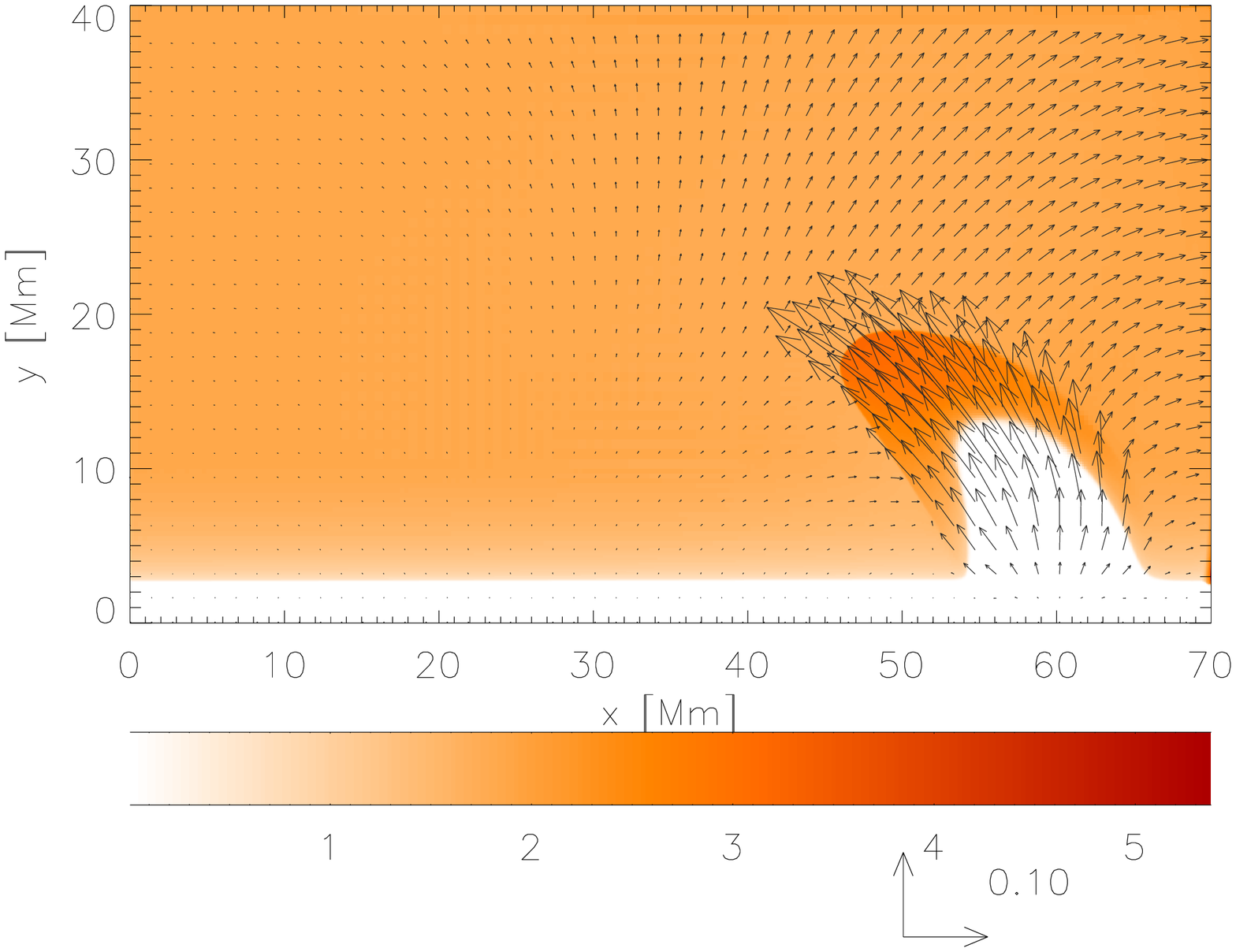}
\includegraphics[scale=0.21, angle=90]{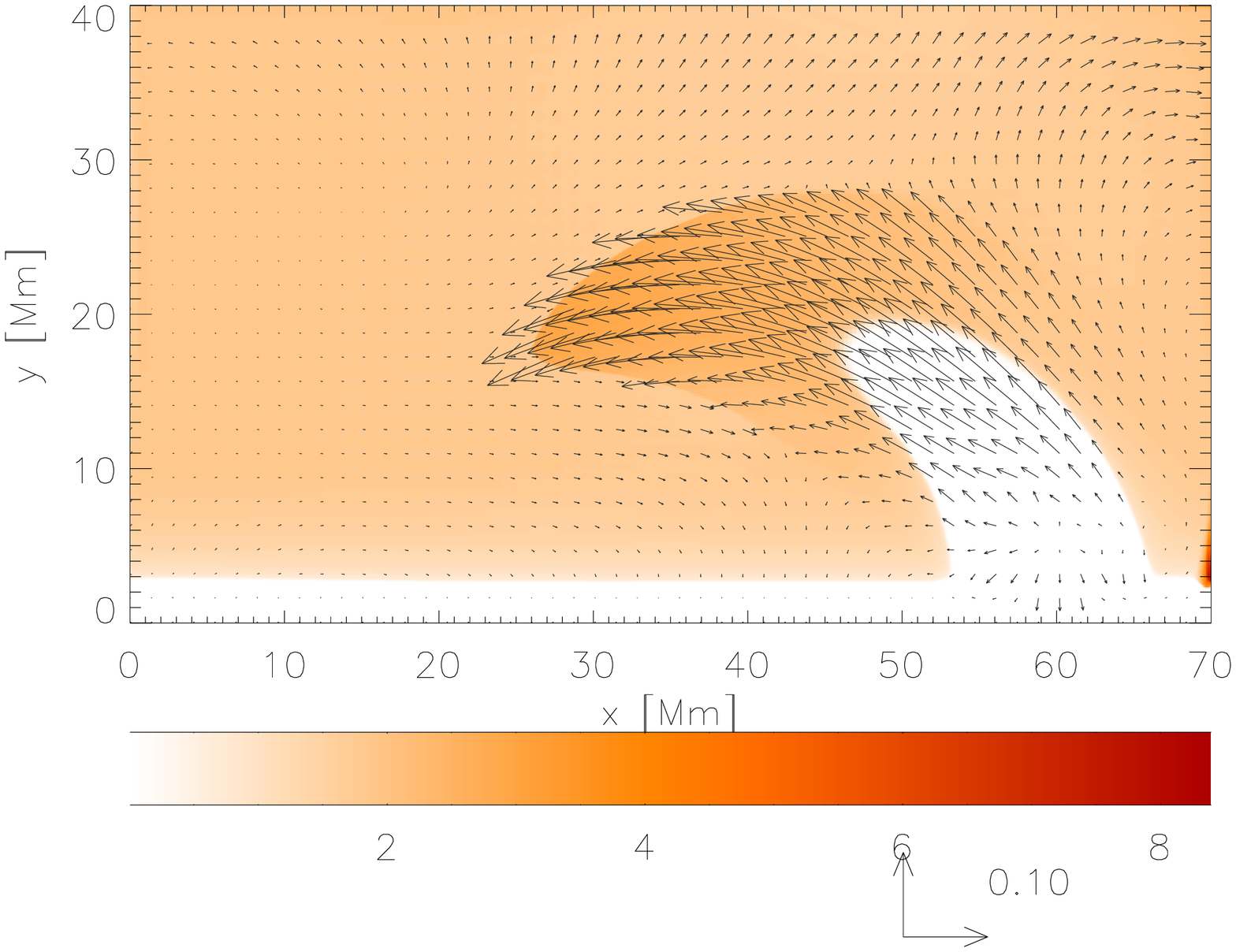}\\
\includegraphics[scale=0.21, angle=90]{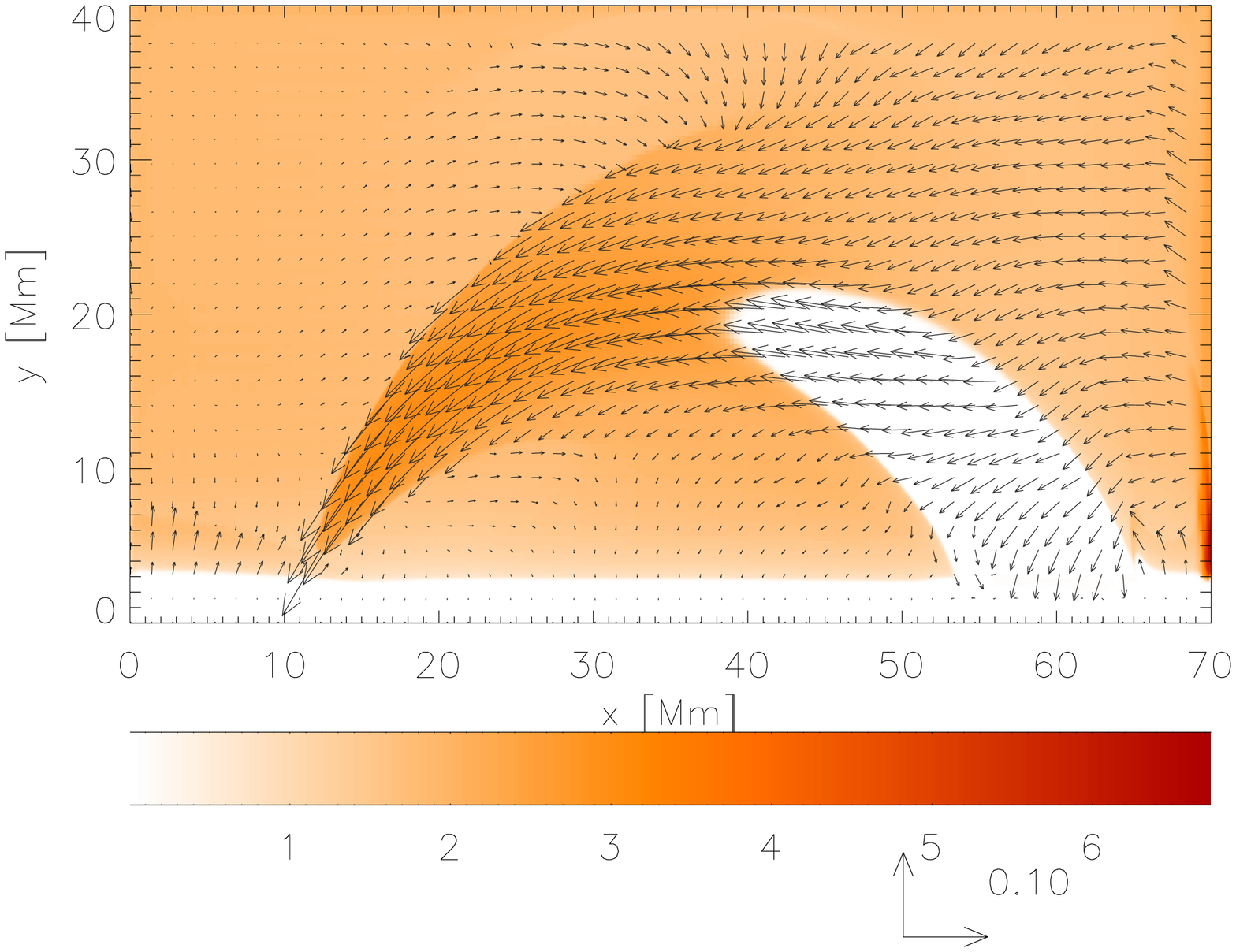}
\includegraphics[scale=0.21, angle=90]{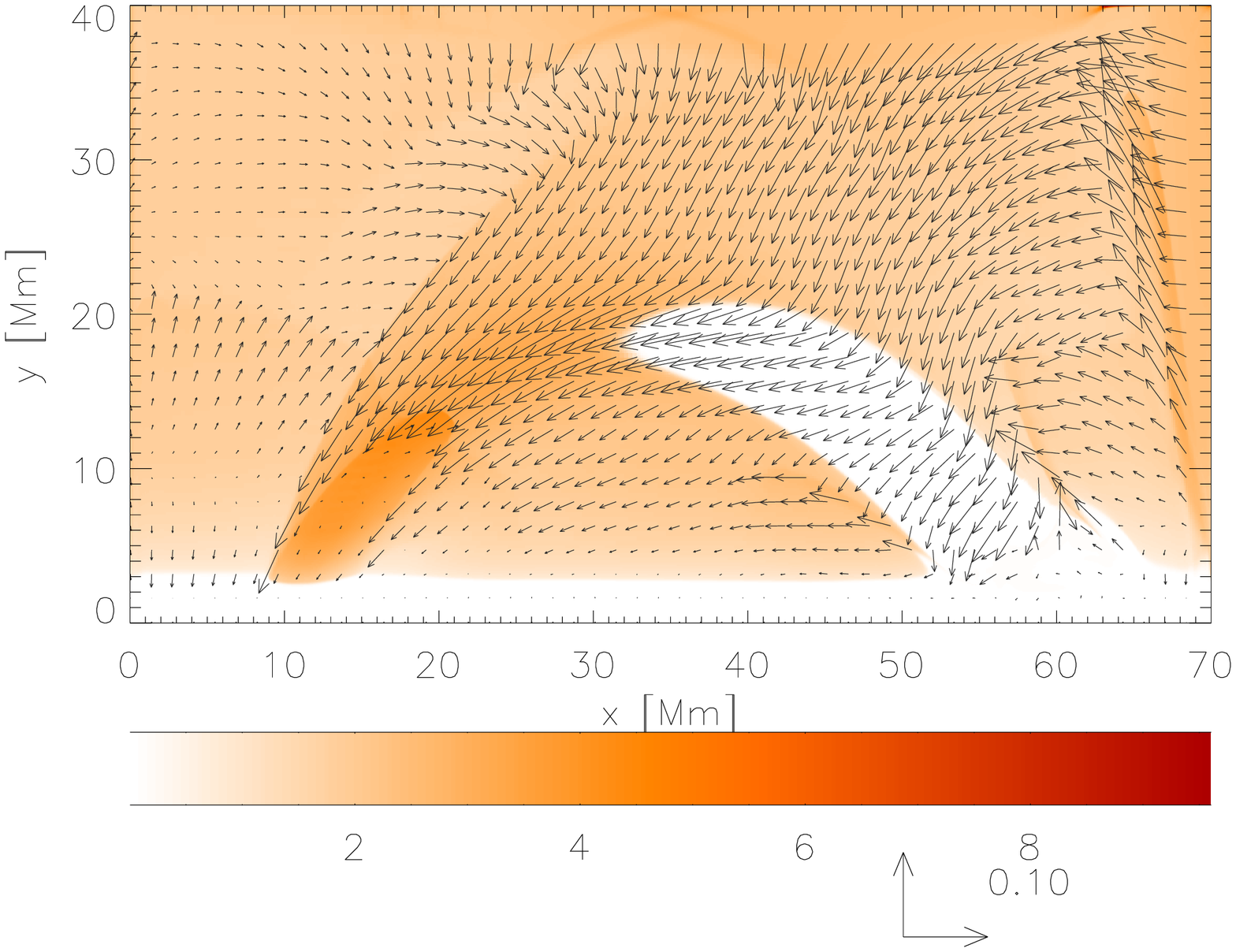}
\includegraphics[scale=0.21, angle=90]{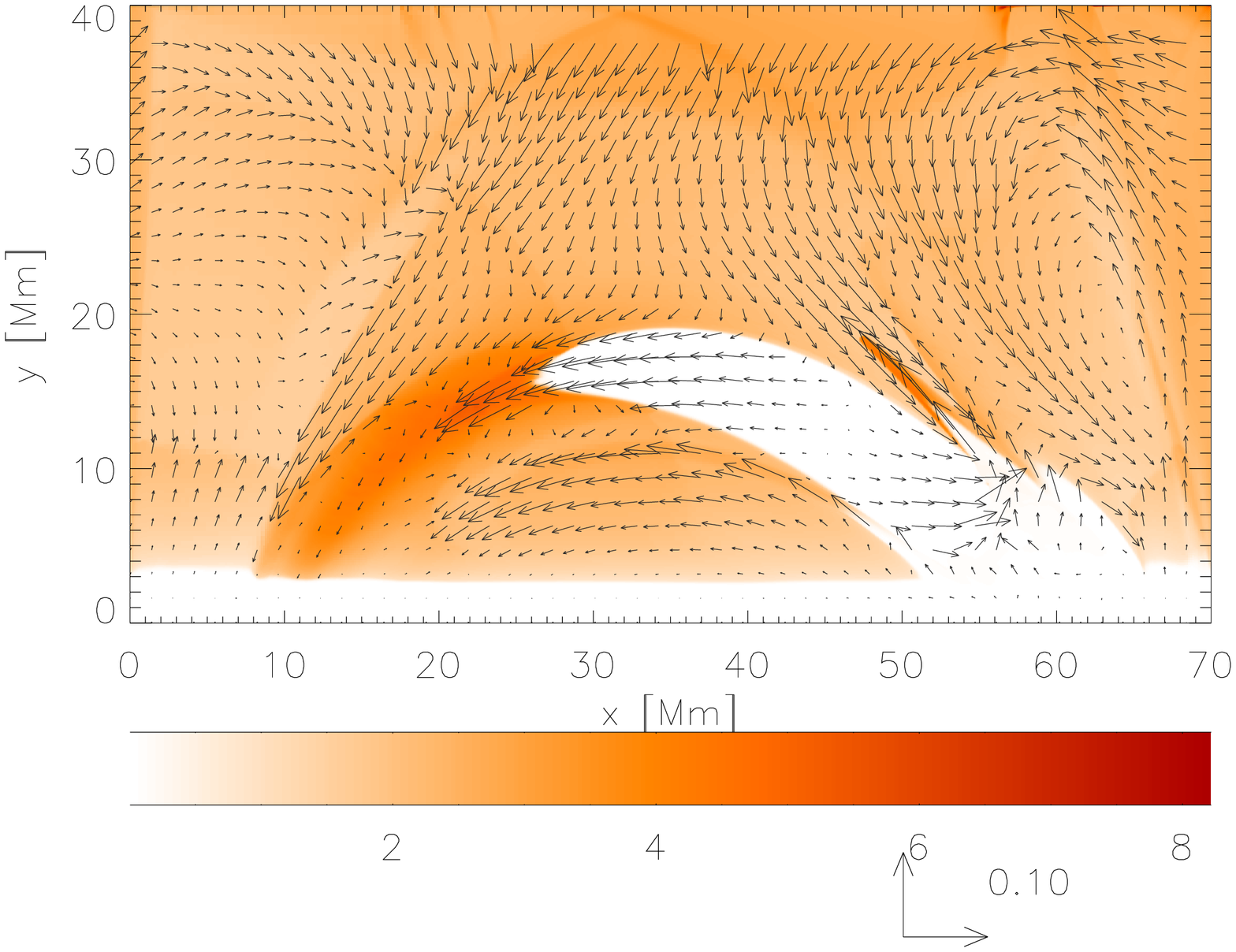}\\
\includegraphics[scale=0.21, angle=90]{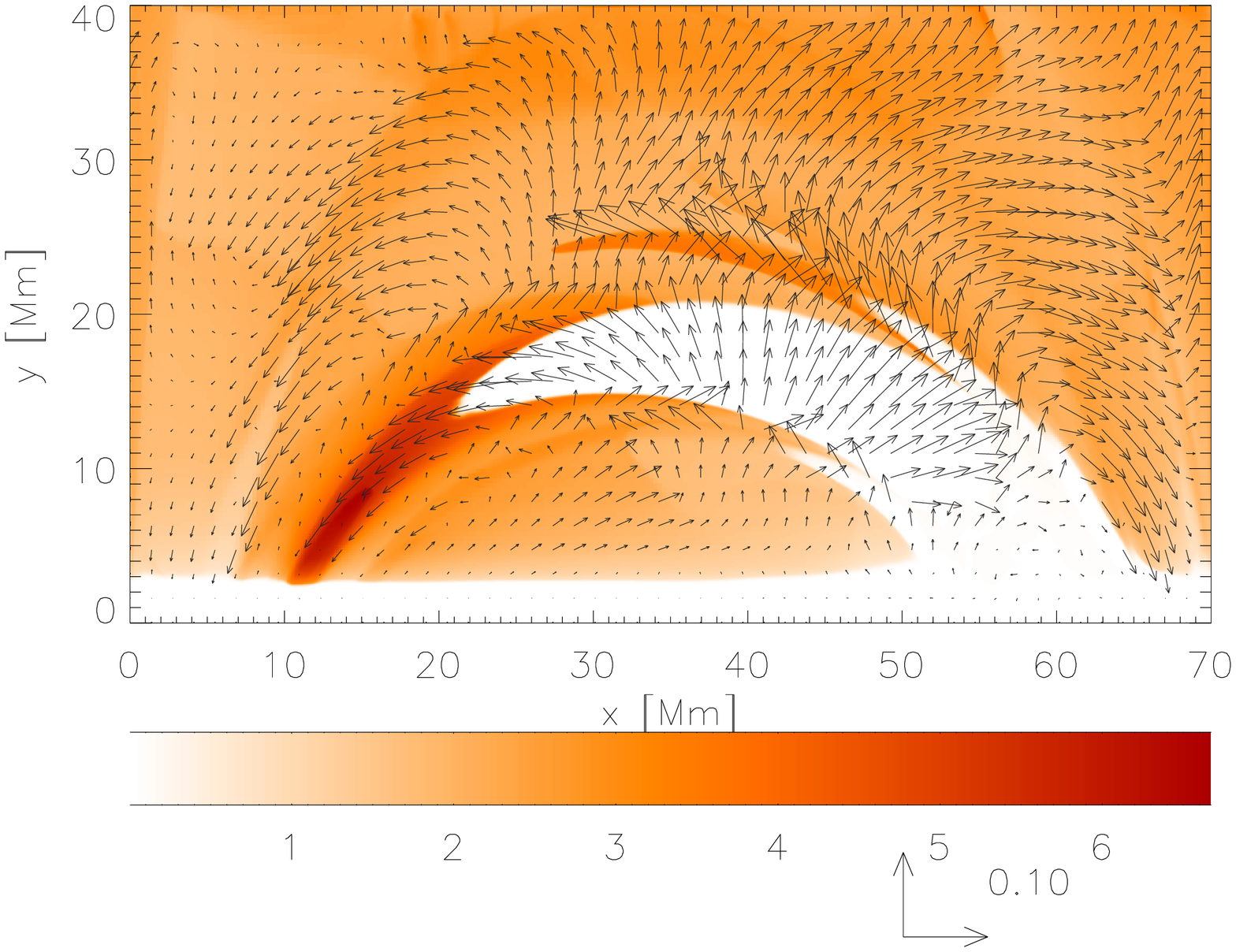}
\includegraphics[scale=0.21, angle=90]{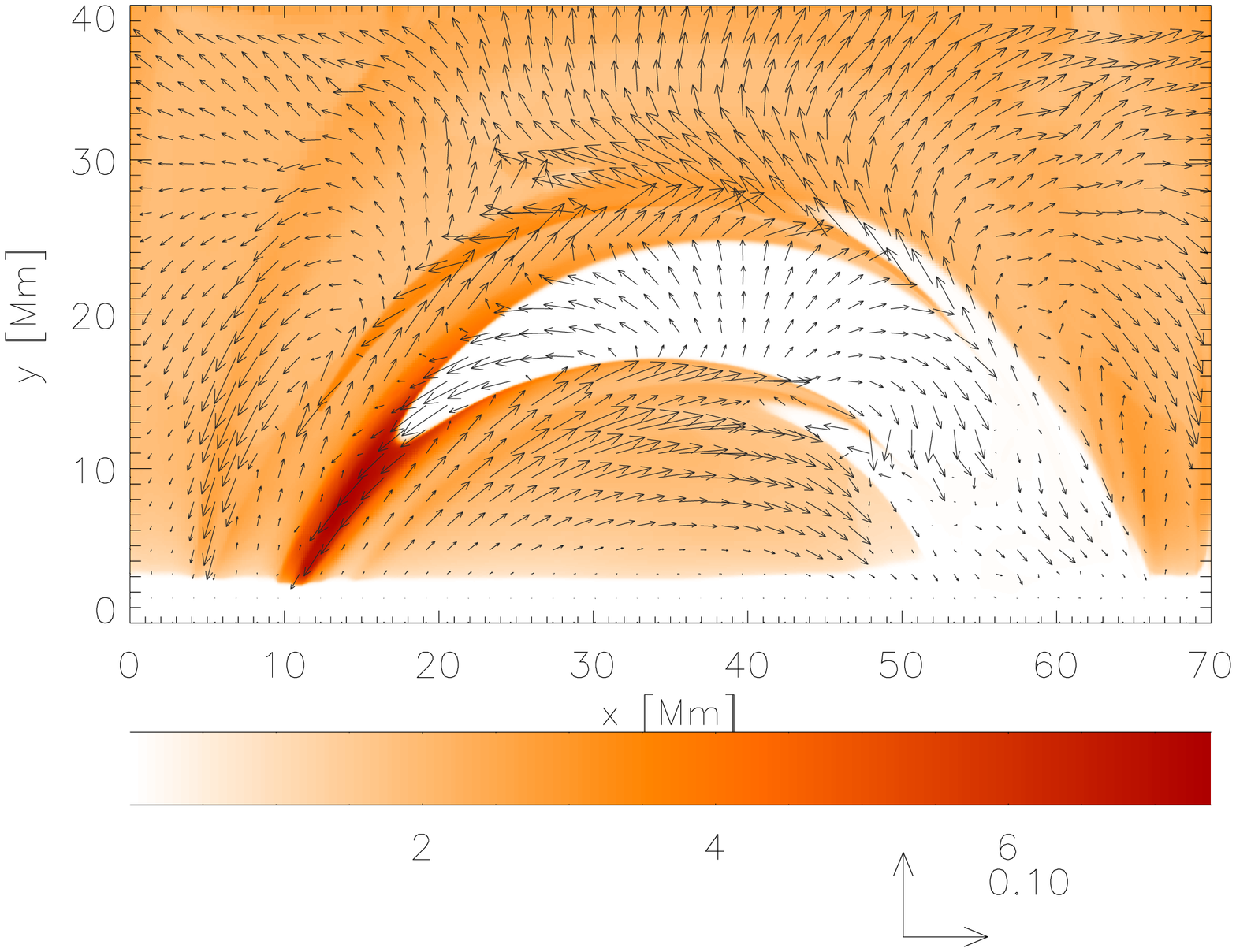}
\includegraphics[scale=0.21, angle=90]{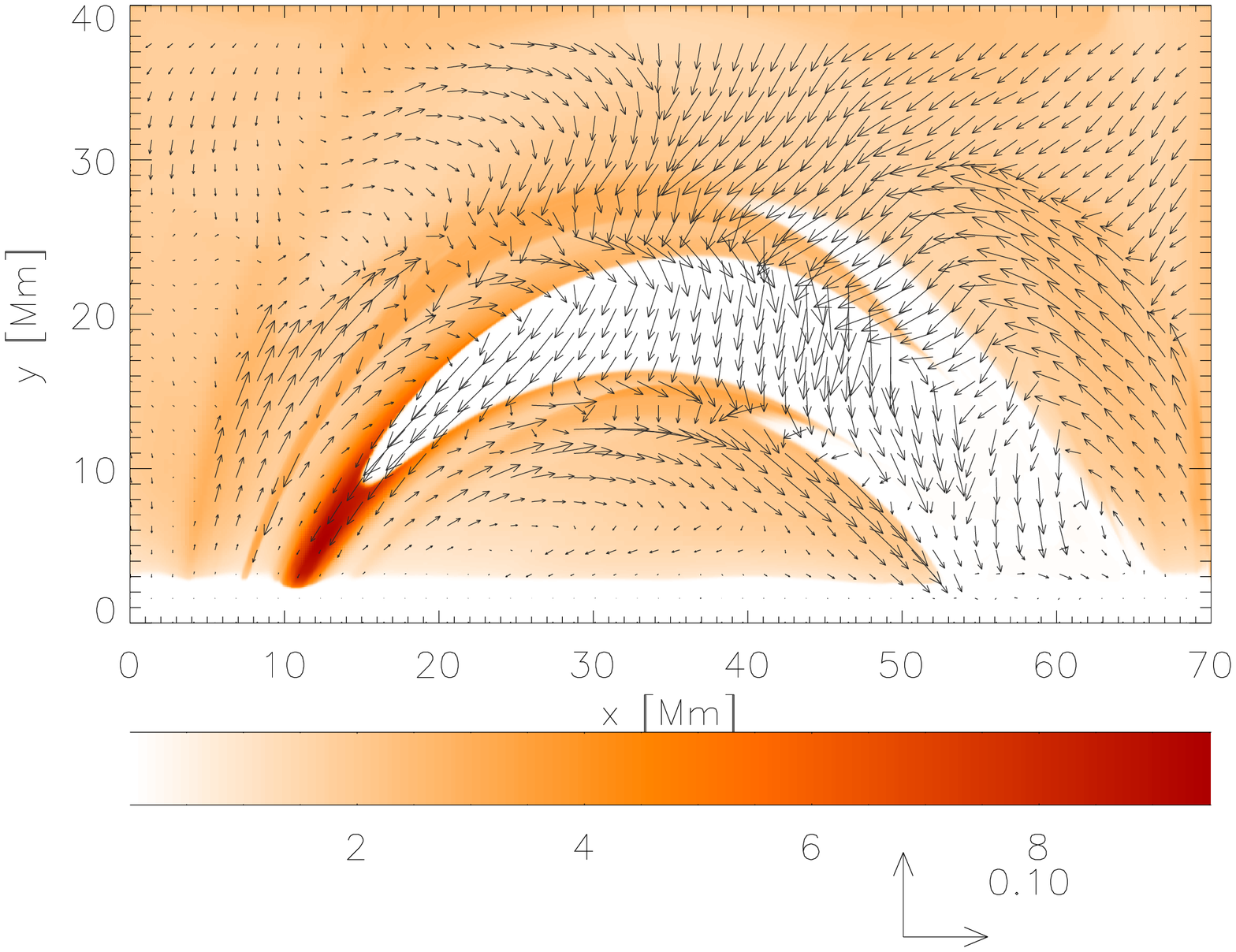}\\
\caption{\small 
Results of numerical simulation : Temperature (colour maps) and velocity (arrows) profiles at
$t=0$ s, $t=100$ s, $t=200$ s, 
$t=300$ s, $t=400$ s, $t=500$ s, 
$t=600$ s, $t=700$ s, and $t=800$ s
(from top-left to bottom-right). 
Temperature is drawn in units of $1$ MK. 
The arrow below each panel represents the length of the velocity vector, expressed in units of $150$ km s$^{-1}$. 
}
\label{fig:jet_prof}
\end{figure*}






\clearpage

\end{document}